# Accuracy of the Water Vapour Content Measurements in the Atmosphere Using Optical Methods

VIATCHESLAV GALKIN [1)], ULRICH LEITERER [2)], GALINA ALEKSEEVA [1)], VICTOR NOVIKOV [1)], AND VSEVOLOD PAKHOMOV [1)]

1) Russian Academy of Sciences, Pulkovo Observatory, 196140 St. Petersburg, Russia
2) German Weather Service, Meteorological Observatory, 15848 Tauche OT Lindenberg, Germany

E-mail: galkin@gao.spb.ru;
Ulrich.Leiterer@dwd.de ;
alekseeva@gao.spb.ru ;
novikov_victor@.mail.ru , novikov@gao.spb.ru

**Summary**

This paper describes the accuracy and the errors of water vapour content measurements in the atmosphere using optical methods, especially starphotometer.

After the general explanations of the used expressions for the star-magnitude observations of the water vapour absorption in section 3 the absorption model for the water vapour band will be discussed. Sections 4 and 5 give an overview on the technique to determine the model parameters both from spectroscopic laboratory and radiosonde observation data.

Finally, the sections 6 and 7 are dealing with the details of the errors; that means errors of observable magnitude, of instrumental extraterrestrial magnitude, of atmospheric extinction determination and of water vapour content determination by radiosonde humidity measurements.

The main conclusion is: Because of the high precision of the results the optical methods for water vapour observation are suited to validate and calibrate alternative methods (GPS, LIDAR, MICROWAVE) which are making constant progress world-wide in these days.

## *1. Introduction*

How the atmospheric water vapour content can be determined by optical methods using an advanced star- and sunphotometer system is described in our paper LEITERER et al., 1998, but a detailed error analysis of the starphotometer method is lacking up to now.
The errors and quality assurance of Lindenberg sunphotometer system are discussed some years ago, e. g. LEITERER and WELLER, 1993 and LEITERER et al., 1994.
The methodical principles of star- and sunphotometry to measure atmospheric constituents (aerosol, water vapour, $O_3$, $NO_2$) have been known for many years. A list of references one can find in the paper of LEITERER et al., 1998 too.

Now we have collected a considerable volume of data using the Lindenberg star- and sunphotometer measuring system. On the results of a part of these data series are reported by NOVIKOV et al., 2000.

Besides the presentation of the results one has to analyse the different errors of the optical methods, especially if a starphotometer is used for water vapour content observation. Night observations of water vapour column content by starphotometer are unusual up to now.
Our aim here is to show the validity of the water vapour column estimation by use of fixed stars light sources and to demonstrate the application to intercomparisons of MICROWAVE related methods.

## *2. General expressions for the star magnitude of the water vapour absorption*

The optical method allows to determine the water vapour content through the relation between the "absorption magnitude" in the water vapour band and the water vapour content on the line of view. The "absorption magnitude" can be obtained by photometric or spectrophotometric observations of stars (GALKIN and ARKHAROV, 1981) on the base of the Equation:

$$\mathbf{m_\lambda^{obs.} = m_\lambda^{o} + \alpha_\lambda \cdot F(z) + \Delta m\ (W)} \qquad (1)$$

where

| | |
|---|---|
| $\mathbf{m_\lambda^{obs.}}$ | - observed star magnitude |
| $\mathbf{m_\lambda^{o}}$ | - extraterrestrial instrumental star magnitude |
| $\alpha_\lambda$ | - extinction for air mass **one** |
| **F(z)** | - air mass |
| **Δm (W)** | - water vapour absorption in star magnitudes |

**W** - water vapour content on the line of view (in **cmppw**)

The coefficient of extinction $\alpha_\lambda$ in the Equation **(1)** includes continuous components of extinction and equals:

$$\alpha_\lambda = \alpha_\lambda^{Ray.} + \alpha_\lambda^{aer.} + \alpha_\lambda^{ozone} \quad (2)$$

where $\alpha_\lambda^{Ray.}$, $\alpha_\lambda^{aer.}$, $\alpha_\lambda^{ozone}$ denote the Rayleigh, aerosol and ozone components of extinction. It is not possible to determine the coefficient of extinction $\alpha_\lambda$ inside the water vapour band directly and therefore $\alpha_\lambda$ must be interpolated to the water vapour wavelength from a region beyond the water vapour band as it is shown in **Figure 1**. We used 0.86 µm and 1.04 µm filters in order to get the interpolated value $\alpha_\lambda$ for $\alpha$ = 0.95 µm water vapour filter and consequently this value $\alpha_\lambda$ depends on extinction errors for those wavelengths and on the interpolation method, too. Thus from Equation **(1)** we derive the next expression for the determination of the "absorption magnitude" $\Delta m(W)$ in the water vapour band:

$$\Delta \mathbf{m}\ (\mathbf{W}) = \mathbf{m}_\lambda^{obs} - \mathbf{m}_\lambda^{o} - \alpha_\lambda \cdot \mathbf{F(z)} \quad (3)$$

Therefore, in order to obtain the absorption magnitude in the water vapour band **$\Delta$m (W)** we must know the observed star magnitude in the "water filter" $\mathbf{m}_\lambda^{obs.}$, the extraterrestrial star magnitude $\mathbf{m}_\lambda^{o}$ for this filter, and the extinction coefficient $\alpha_\lambda$, interpolated between another filters' measurements without water vapour absorption.

### 3. *Model of absorption in the water vapour band*

It is necessary to use an absorption model which connects the absorption magnitude **$\Delta$m (W)** with the water vapour content on the line of view. We used an empirical model developed by GOLUBITSKY and MOSKALENKO (1968) and GALKIN and ARKHAROV (1981):

$$\Delta \mathbf{m}\ (\mathbf{W}) = \mathbf{C}_\lambda \cdot [\mathbf{W}_o \cdot \mathbf{F(z)}]^{\mu_\lambda} \quad (4)$$

where $\mathbf{C}_\lambda$ and $\mu_\lambda$ are empirical parameters, $\mathbf{W}_o$ – the water vapour content in the atmospheric vertical column. We investigated the validity of this model and determined several empirical parameters (ALEKSEEVA et al., 1994). According to our data the Equation **(4)** comes true very well for water vapour contents from 0.2 cmppw to 3.5 cmppw and for pressures from 0.1 atm to 1.0 atm. The Equation **(4)** is fulfilled for the given pressure, and we found out that for different pressures empirical parameter $\mathbf{C}_\lambda$ changes as follows:

$$\mathbf{C}_\lambda(\mathbf{P}) = \mathbf{C}_\lambda(\mathbf{P=1atm}) \cdot \mathbf{P}^{\,n} \quad (5)$$

where **P** is given in **atm**. For the "water filter" used in the Lindenberg starphotometer one gets **n** = 0.44.

The pressure in the atmosphere changes a according to the pressure from 1 to 0 atm and the water vapour is distributed in a very complicated manner. Taking into account a water vapour distribution with the height W(h), we can determine the effective pressure $\mathbf{P_{eff.}}$ for water vapour in the atmosphere as:

$$\mathbf{P_{eff.} = \int P(h) \cdot W(h)dh \,/ \int W(h)dh} \qquad \mathbf{(6)}$$

where **P(h)** is the change of pressure with the altitude. If on sea level the pressure is equal to $\mathbf{P_o}$, we get $\mathbf{P_{eff.}} = 0.845 \cdot \mathbf{P_o}$ for the mean distribution of water vapour in the atmosphere of the "Lindenberg Column". Because pressure influences the quantity of water vapour absorption according to the half-width and shape of the absorption coefficient of individual lines, it is very important that the shapes of the water vapour telluric lines are Lorenz shapes. This circumstance allows us to make a very simply transfer from a homogeneous layer of water vapour in the laboratory to a non-homogeneous atmosphere by means of half-width of the telluric line $\gamma_{tel.}$, which is equal:

$$\gamma_{tel.} = \gamma_{eff.} = \mathbf{P_{eff.}} \cdot \gamma_o \qquad \mathbf{(7)}$$

where $\gamma_o$ is the half-width of water vapour line for $\mathbf{P} = 1$ atm. Now we obtain the empirical parameter $\mathbf{C_\lambda}$ for the non-homogeneous atmosphere taking it for $\mathbf{P_{eff.}} = 0.845 \cdot \mathbf{P_o}$. Of course the real $\mathbf{P_{eff.}}$ can change from day to day. That may lead to a tolerance of a few percent, but the manner by which one may introduce the correlative correction will be discussed at another place.

Thus for the Lindenberg "water filter" $\lambda = 0.946$ µm (HBW = 0.0068µm) with an assumed "ideal" rectangular shape one gets:

$$\mathbf{C_\lambda = 0.732 \qquad \mu_\lambda = 0.594} \qquad \mathbf{(8)}$$

It is necessary to remark that the atmosphere is not only non-homogeneous but also non-isothermal. The absorption dependence on temperature can be introduced by means of the parameter $\mathbf{C_\lambda}$ as well, but we do not have any data for this procedure so far. Based on general consideration we may suppose that the temperature influence is not big. There are two reasons for the temperature variation's influence on absorption. Firstly, the number of absorbing molecules on a given energy level depends on temperature. But in the water band the distribution of absorption lines along the spectrum is accidental, and for every wavelength the principal part of absorption is determined by strong lines forming from energy levels on which the number of absorbing molecules depends only slightly on temperature. The next reason for the temperature influence on absorption is the dependence of the half-width of a line on temperature. But again, the absorption is proportional to the square root of the half-width of the line, and the half-width is proportional to the square root of the absolute temperature. Therefore the absorption depends on the temperature to the power of a quarter. Detail analysis of this dependence can be completed both theoretically and experimentally, for example by determination of the empirical parameter $\mathbf{C_\lambda}$ in summer and in winter. Without taking into account the temperature, the parameters of Expression **(8)** correspond to the homogeneous and isothermal atmosphere with a temperature of $+15^{O}C$ for which the laboratory empirical parameters were obtained.

## 4. *Calculation of empirical parameters from the laboratory data*

The parameters $\mathbf{C_\lambda}$, $\boldsymbol{\mu_\lambda}$ (see Expression **8**) correspond to a "rectangular signal" of a "ideal" filter which is absorbed by water vapour.

For a "real" filter, see **Figure 2**, the light transmission ca be described by the Equation:

$$\mathbf{T(W) = \int T_\lambda(W) \cdot S_\lambda \cdot E_\lambda \cdot R_\lambda d\lambda \, / \int S_\lambda \cdot E_\lambda \cdot R_\lambda d\lambda} \qquad \textbf{(9)}$$

where

| | |
|---|---|
| **T(W)** | - transmission of the signal which is formed by the properties of filter and receiver |
| $\mathbf{T_\lambda(W)}$ | - spectral transmission of water vapour |
| $\mathbf{S_\lambda}$ | - spectral curve of filter transmission |
| $\mathbf{E_\lambda}$ | - spectral energy distribution in the source of radiation |
| $\mathbf{R_\lambda}$ | - spectral curve of receiver sensitivity |

Using Equation **(9)** we can calculate **T(W)** for different **W** and then translate it to the star magnitude using the expression:

$$\mathbf{\Delta m\ (W) = -2.5 \cdot log\ T(W)} \qquad \textbf{(10)}$$

The dependence $\mathbf{\Delta m\ (W)}$ on water vapour content may be approximated by a power function, see Equation **(4)**, to get the empirical parameters $\mathbf{C_\lambda}$ and $\boldsymbol{\mu_\lambda}$:

$$\mathbf{\Delta m\ (W) = C_\lambda \cdot W^{\mu_\lambda}} \qquad \textbf{(11)}$$

The empirical parameters $\mathbf{C_\lambda}$ and $\boldsymbol{\mu_\lambda}$ obtained in this way for the "water filter" ($\lambda$ = 0.946 µm, HBW = 0.0068µm) of the star photometer in Lindenberg are equal:

$$C_\lambda = 0.589 \quad \text{and} \quad \mu_\lambda = 0,560 \qquad (12)$$

The empirical parameters $C_\lambda$ and $\mu_\lambda$ for the sun photometer were calculated analogously. These parameters were calculated on the basis of laboratory measurements, ALEKSEEVA, 1994, with Pulkovo's multipassage vacuum cuvette VKM-100 (length of line up to 4 km), and allow the independent determination of water vapour contents in the atmosphere.

## *5. Determination of empirical parameters by observation*

It is also possible to obtain empirical parameters $C_\lambda$, $\mu_\lambda$ from observations with star or solar photometers if the content of water vapour in the atmosphere is known from other sources, for example from radiosonde (RS) data. Because it is impossible to obtain extraterrestrial magnitudes $m^o_\lambda$ in the water vapour band (the application of Bouger-Lambert Law delivers uncorrected $m^o_\lambda$), we used the model spectrophotometric value $\mu_\lambda = 0.594$ (Expression **8)** as a first approximation. Then the magnitudes $m^o_\lambda$, $\Delta m\ (W)$ and $C_\lambda$ were calculated applying Equation **(3), (4)**. Obtained $C_\lambda$ and accepted $\mu_\lambda$ were used to determine again $m^o_\lambda$, $\Delta m\ (W)$ and **Wo**. The starphotometer $W_o$ values were compared with **$W_o$(RS)** values derived from interpolated radiosonde humidity profiles (RS-data), and power function approximation (Equation **(4)**) of $\Delta m\ (W)$ was applied to determine the parameters $C_\lambda$ and $\mu_\lambda$ again. If the new calculated parameters are different from the accepted parameters for calculating $m^o$ and **Wo**, one can repeat the procedure of $C_\lambda$, $\mu_\lambda$-parameter determination to get a "optimal" agreement between **Wo** and **Wo(RS)**. The best parameters for observation in 1997 were

$$C_\lambda = 0.598 \quad \text{and} \quad \mu_\lambda = 0,564 \qquad \text{(Oct.-Nov. 1997)} \qquad (13)$$

It is necessary to remark that there are many combinations $C_\lambda$, $\mu_\lambda$ and $m^o$ which give a good representation of observation data. From these combinations we used the combination which shows the best agreement between **Wo** and interpolated **$W_o$(RS)**-data **(13)**.

The observations of the years 1998 and 1999 were calculated with $C_\lambda$, $\mu_\lambda$-parameters **(13)** also. Using parameters (13) we remained in the same system of the instrumental extraterrestrial magnitudes $m^o$, but the approximation $\Delta m\ (W)$ by the power function (Equation **(4)**) lead to other parameters $C_\lambda$ and $\mu_\lambda$ for the different periods of observations.

So the best parameters for the representation of observation data for the periods 1998 and 1999 were:

$$\mathbf{C_\lambda = 0.^m582 \quad and \quad \mu_\lambda = 0{,}548 \quad (June\text{-}Aug.\ 1998)} \qquad \mathbf{(14)}$$

$$\mathbf{C_\lambda = 0.^m588 \quad and \quad \mu_\lambda = 0{,}553 \quad (Sep.\text{-}Oct.\ 1999)} \qquad \mathbf{(15)}$$

One can see some differences between the values of these parameters $\mathbf{C_\lambda}$, $\boldsymbol{\mu_\lambda}$ in the different years and seasons. We have different temperatures and $P_{eff.}$ for various seasons of observations and we must take into consideration the influence of these parameters on absorption in the process of calculations. Moreover there are accidental and systematic errors of $\mathbf{\Delta m\ (W)}$ determination. We have only one $\mathbf{W_o(RS)}$-measurement every 6 hours. And we are compelled to use a linear interpolation of $\mathbf{W_o(RS)}$-data through this interval for approximation of $\mathbf{\Delta m\ (W)}$ by the power function. This procedure may be statistically correct only for large volumes of data. In **Figure 3** the example of the approximation $\mathbf{\Delta m\ (W)}$ by the power function for observations in 1999 (parameters **(15)**) is shown, and in **Figure 4** the comparison **Wo** with interpolated $\mathbf{W_o(RS)}$-data is given for the same observations, with

$$\mathbf{W(RS) = Wo(RS) \cdot F(z)} \qquad \mathbf{(16)}$$

denotes the water vapour content on the line of view with airmass **F(z)**, computed from vertical radiosonde humidity profile derived water vapour content **Wo(RS).** The small systematic error in $\mathbf{W_o/Wo(RS)}$ in **Figure 4** may be caused by the linear interpolation of **Wo(RS)**-data (the volume of observable data is not so large).

## 6. *Determination errors of the absorption magnitude of water vapour*

In order to improve the accuracy of the water vapour content determination it is necessary to reveal and to remove several sources of errors. Every part of expression **(3)** has its own error: the error of the observed magnitude $\mathbf{\delta m_\lambda^{obs.}}$, the error of the instrumental extraterrestrial magnitude $\mathbf{\delta m_\lambda^{o}}$, the error of the extinction determination $\delta\alpha_\lambda$ and the error of the water vapour absorption magnitude $\mathbf{\delta m_\lambda^{w}}$. For us it is most important to know $\mathbf{\delta m_\lambda^{w}}$, which can be determined as follows:

$$\mathbf{(\delta m_\lambda^{w})^2 = (\delta m_\lambda^{obs.})^2 + (\delta m_\lambda^{o})^2 + (\delta\alpha_\lambda)^2 \cdot F^2(z)} \qquad \mathbf{(17)}$$

What are the sources of these errors?

The error of observation $\delta \mathbf{m}_\lambda^{obs.}$ depends on the stability of the detector, the magnitude of the input signal and on the atmospheric scintillation. The error of the extraterrestrial magnitude $\delta \mathbf{m}_\lambda^{o}$ depends on the error of the observation $\delta \mathbf{m}_\lambda^{obs.}$, the long-time stability of the receiver, and on the error of the extinction determination. The latter is connected to the accuracy of the magnitudes $\mathbf{m}_\lambda^{obs.}$ and $\mathbf{m}_\lambda^{o}$, and the error of the extinction time-dependence used. Moreover, the extinction for the water vapour band wavelength may be received only by interpolation from filters aside from the water vapour band and consequently $\delta\alpha_\lambda$ depends from the way of $\alpha_\lambda$-interpolation also.

Which errors we have today?

In **Figure 5** the errors $\delta \mathbf{m}_\lambda^{obs}$ in dependence on the observed star magnitudes $\mathbf{m}_\lambda^{obs}$ are shown for two nights, 9th and 13th September 1999. This error is calculated using a number of iterations for the given observation . Because the time of the data accumulation is different for every filter and every observation, **Figure 5** presents the statistical character of the distribution and tendency of the errors. There are two curves in **Figure 5**. The lower curve corresponds to high stars which have **F(z)** = 1.0-1.2 and the second upper curve corresponds to lower stars with **F(z)** = 2 - 3. Of course, the values of the errors depend on the signal magnitude, but for bright stars (negative magnitudes) we see a limit, after which the errors are not diminished with growing brightness. This limit is connected to the atmospheric scintillation. For the given telescope, detector, and for the accepted conditions of observation we consequently get this maximum accuracy of observation:

$$\delta \mathbf{m}_\lambda^{obs.} = 0.^m005 \text{ resp. } 0.5\% \text{ (for high stars)}$$

**(18)**

$$\delta \mathbf{m}_\lambda^{obs.} = 0.^m015 \text{ resp. } 1.5\% \text{(for low stars)}$$

This is the our best accuracy today. How do you conserve the accuracy obtained for one observation during the night? In **Figure 6** the deviations of single high star measurements $\mathbf{m}_\lambda^{obs}$ to the mean trend curve of all 111 measurements are shown. One can see that the observation error is conserved during a measuring period of 3 to 4 hours. So we assume that the real error of observation for the high star is nearly $0.^m005$ (0.5%).

What is the individual instrumental extraterrestrial star magnitude error $\delta \mathbf{m}_\lambda^{o}$? In **Figure 7a** the error $\delta \mathbf{m}_\lambda^{o}$ is shown for individual observations of high and low stars in dependence on extraterrestrial star magnitudes applying the “Two Star Differential Method”, as described by LEITERER et al., 1998. The errors for high and low stars have systematic differences as follows:

$$\delta m_\lambda^o = 0.^m04 \text{ resp. } 4.0\% \text{ (for high stars)}$$

**(19)**

$$\delta m_\lambda^o = 0.^m 10 \text{esp. } 10.0\% \text{ (for low stars)}$$

It is clear that we must not use low stars for determinations of $\mathbf{m_\lambda^o}$, but sometimes we are forced to do it if we do not have enough high stars in order to determine the photometric system reliably. In **Figure 7b** the distribution of $\boldsymbol{\delta m_\lambda^o}$-errors is shown for mean extraterrestrial magnitudes $\mathbf{m_\lambda^o}$, which we used for the extinction determination after the "Second Approximation" (see LEITERER et al., 1998). Although for the most stars $\boldsymbol{\delta m_\lambda^o}$ is equal to $0.^m01$ and smaller, for a few stars the error $\boldsymbol{\delta m_\lambda^o}$ is larger. That is connected to the use of low stars for mean $\mathbf{m_\lambda^o}$ determination.

There are several reasons for the instability of the photometric system during the whole period of observations (1997-1999):

- bad adjustment of the filter number 2,
- low photometer sensitivity for filters 1 and 10,
- an incorrect optical scheme until the photometer optical system was changed in June 1999,
- the dependence of the photometric system sensibility on temperature.

In any case we can see that the computed individual extraterrestrial $\mathbf{m_\lambda^o}$-errors **(19)** are larger than the observed $\mathbf{m_\lambda^{obs.}}$-errors **(18)**. At what point does the loss of accuracy occur? The analysis of the data reduction procedure shows that using the extinction dependence on time always leads to an error of $\delta\alpha_\lambda \sim 0.^m02 - 0.^m03$ (2-3%).

The main sources of the errors are:

- Real spatial variations of extinction
- Unsuccessful approximation of extinction time-variations
- Determination errors of the extraterrestrial instrumental $\mathbf{m_\lambda^o}$-data by the "Second Approximation"
- Instability of the photometric system (especially due to temperature changes)

A detailed analysis of these problems is necessary, including the determination of revised absolute extraterrestrial magnitudes for the measuring channels.

## 7. *Errors of the water vapour content determination*

After the reconstruction of the starphotometer in June 1999 (see Appendix 1) some error sources are reduced. Results of the error analysis are shown in **Figures 8** to **11**. In **Figure 8** the calculated errors of the water vapour content determination for 13th September 1999 are shown. The values of the errors are ~5-6% both for high and low stars. In **Figure 9** the distributions of observation errors for 5 good nights are shown in dependence on the water vapour content $\mathbf{W_o}$. In **Figure 9** the calculated error-distributions for the values $\delta \mathbf{m}^w$ = $\mathbf{0.^m007}$, $\delta \mathbf{m}^w = \mathbf{0.^m020}$ and $\delta \mathbf{m}^w = \mathbf{0.^m030}$ are also presented. One can see that the distribution of errors corresponds to $\delta \mathbf{m}^w = \mathbf{0.^m025}$. The distribution for $\delta \mathbf{m}^w = \mathbf{0.^m007}$ is the ideal case when principal errors connected to observed magnitudes, errors of extinction, and extraterrestrial magnitudes are small. It is the real situation for a stable instrumental system, stable and homogeneous extinction. Consequently the error **~1%** ($\delta \mathbf{m}^w = \mathbf{0.^m007}$) for the water vapour content is the set goal. Yet today we have cases when errors of the water vapour content determinations are as high as 2-3%, as we can see from comparison with $\mathbf{W_o(RS)}$-data for 9th September 1999 (**Figure 10**). For the longer period (09th to 15th September 1999) the comparison with $\mathbf{W_o(RS)}$-data is given in **Figure 11.** The calculated errors for those data are shown in **Figure 9**. One can see that the comparison of the starphotometer data with external $\mathbf{W_o(RS)}$-data does not contradict the calculated errors. The accuracy of the water vapour content determination is sufficient to investigate the water vapour actual time-variations during the night. The example for such rapid monitoring ($\Delta t \sim 3$ min. between individual star observations) is shown in **Figure 12** for night of August 11/12th 1998 (**LACE** experiment **98**) is shown in **Figure 12** together with radiosonde (interpolated) and microwave radiometer data (mean for $\Delta t = 10$ min).

## *8. Conclusion*

Trying to determine the water vapour content in the atmosphere we are confronted with two problem areas.

The first is the **astronomical** one: how to conduct the observations with the best results, receive instrumental extraterrestrial magnitudes for the „Second Approximation“, determine the extinction, and choose the part of extinction connected to the water vapour absorption at last.

The second is the **spectroscopic** problem, namely how to choose the model of absorption in the water vapour band, which must connect the absorption magnitude with the water vapour content on the line of view. We have to investigate how the parameters of absorption $\mathbf{C_\lambda}$ and $\mu_\lambda$ are changed due to the variations of atmospheric parameters - temperature, pressure and distribution of the water vapour depending on the height. Also it is necessary to choose parameters of the model and to determine them in the best way.

In this report we examined all these problems and gave typical magnitudes for different parameters and their errors on the base of observations with the star photometer in Lindenberg. Today the precision of the water vapour content determinations by optical methods reaches 2-3% and it could reach 1% and better after all steps of observations and determination of the empirical parameters and star magnitudes have been improved. The very important property of the optical method is the independence of calibration which can be obtained from the laboratory data. The independence of calibration and high precision make the optical method very useful for calibration and control of other methods (MICROWAVE, LIDAR, GPS) to determine the water vapour content in the atmosphere.

***Acknowledgements***

This work was funded by the Russian-German project of the Deutsche Forschungsgemeinschaft (DFG) as project 436 RUS 133/76 and Russian Foundation for Basic Research (RFBR) as project 01-05-04000 ННИО_a. We are grateful to Mr. Dietmar Dauß and Mrs. Tatjana Naebert who both have given technical assistance. We like also to express our thanks to Ms. Kathleen Dix who prepared the files for text and figures.

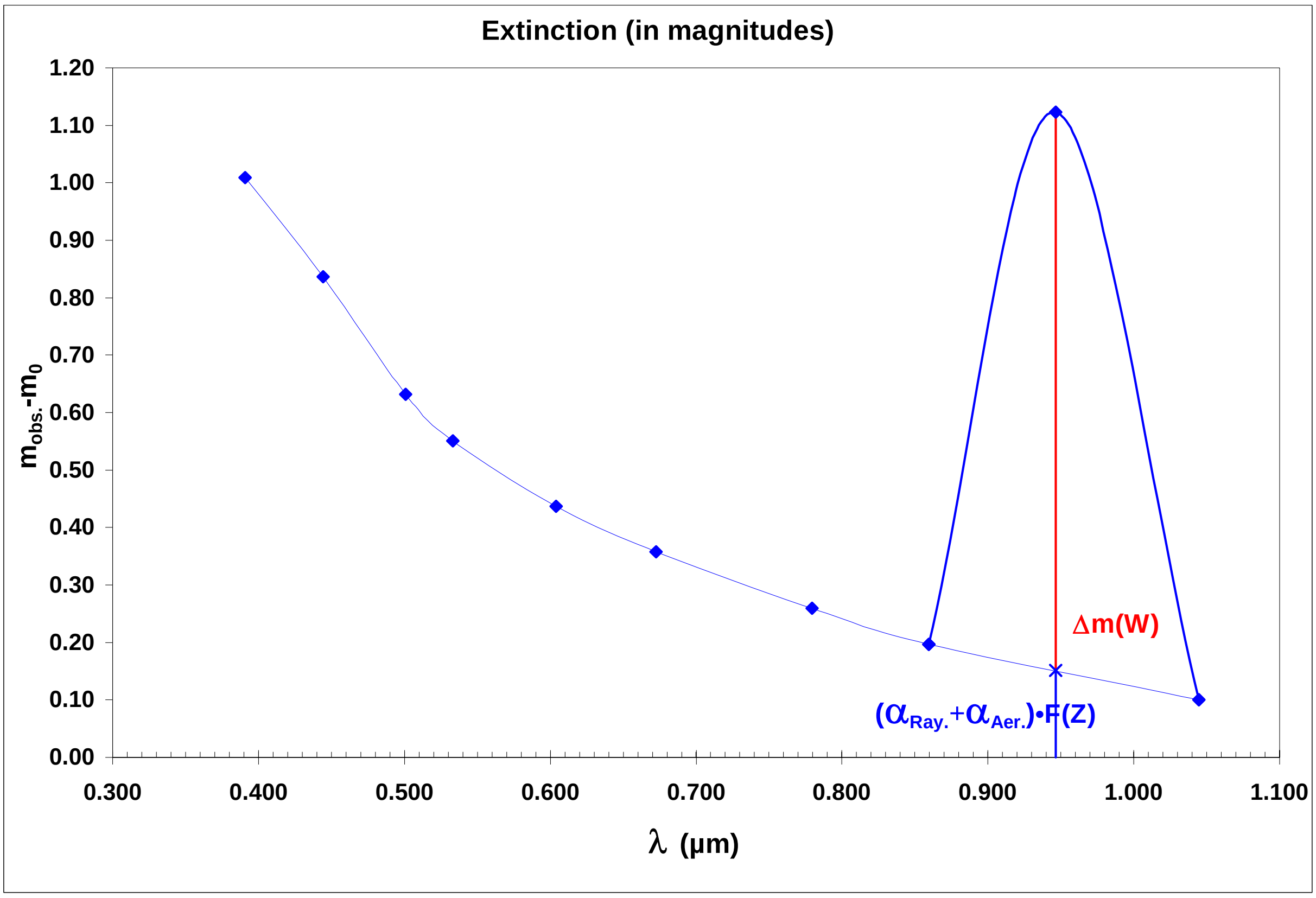

**Figure 1.** Extinction as function from wavelength.

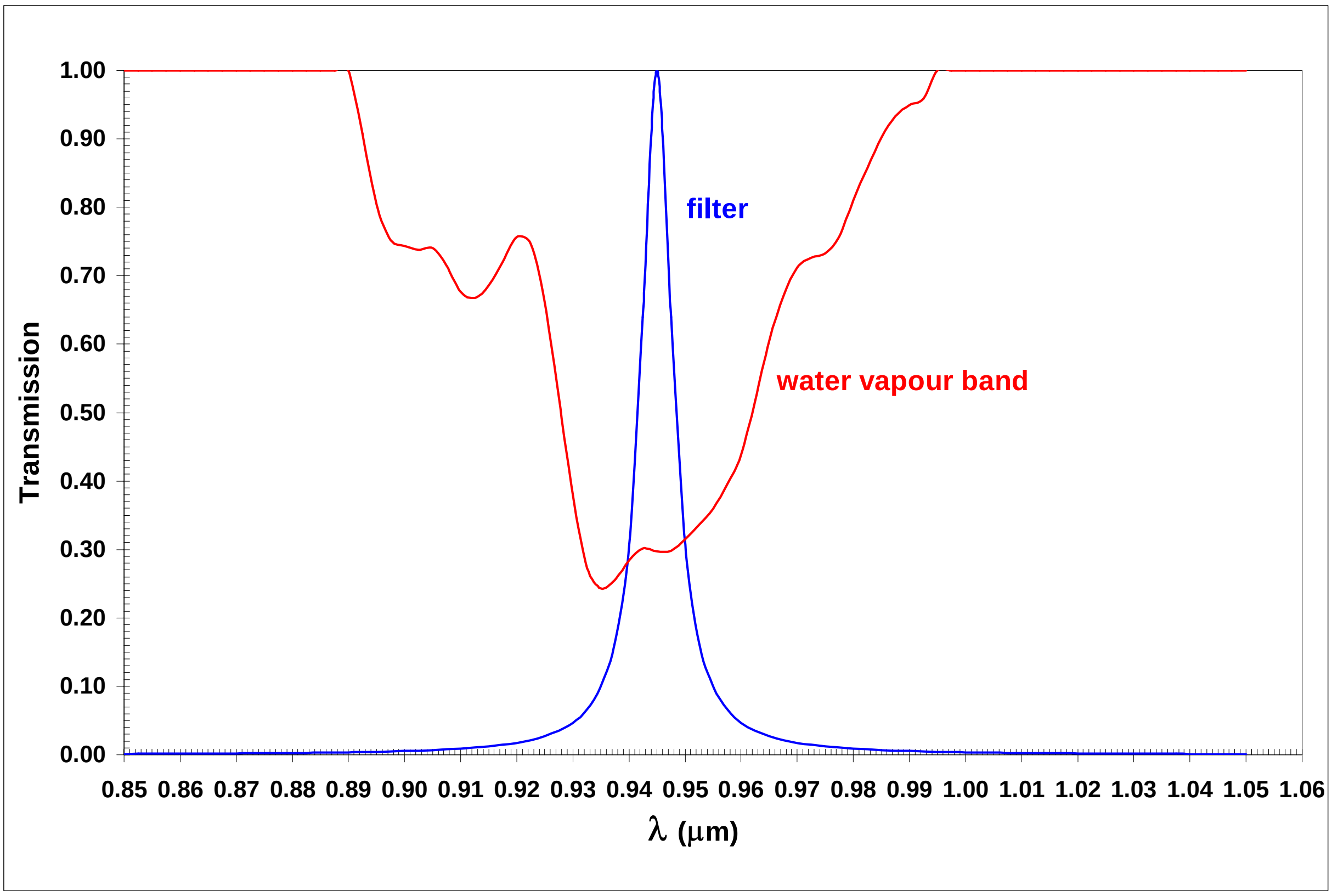

**Figure 2.** Transmission of water vapour band and “water” filter.

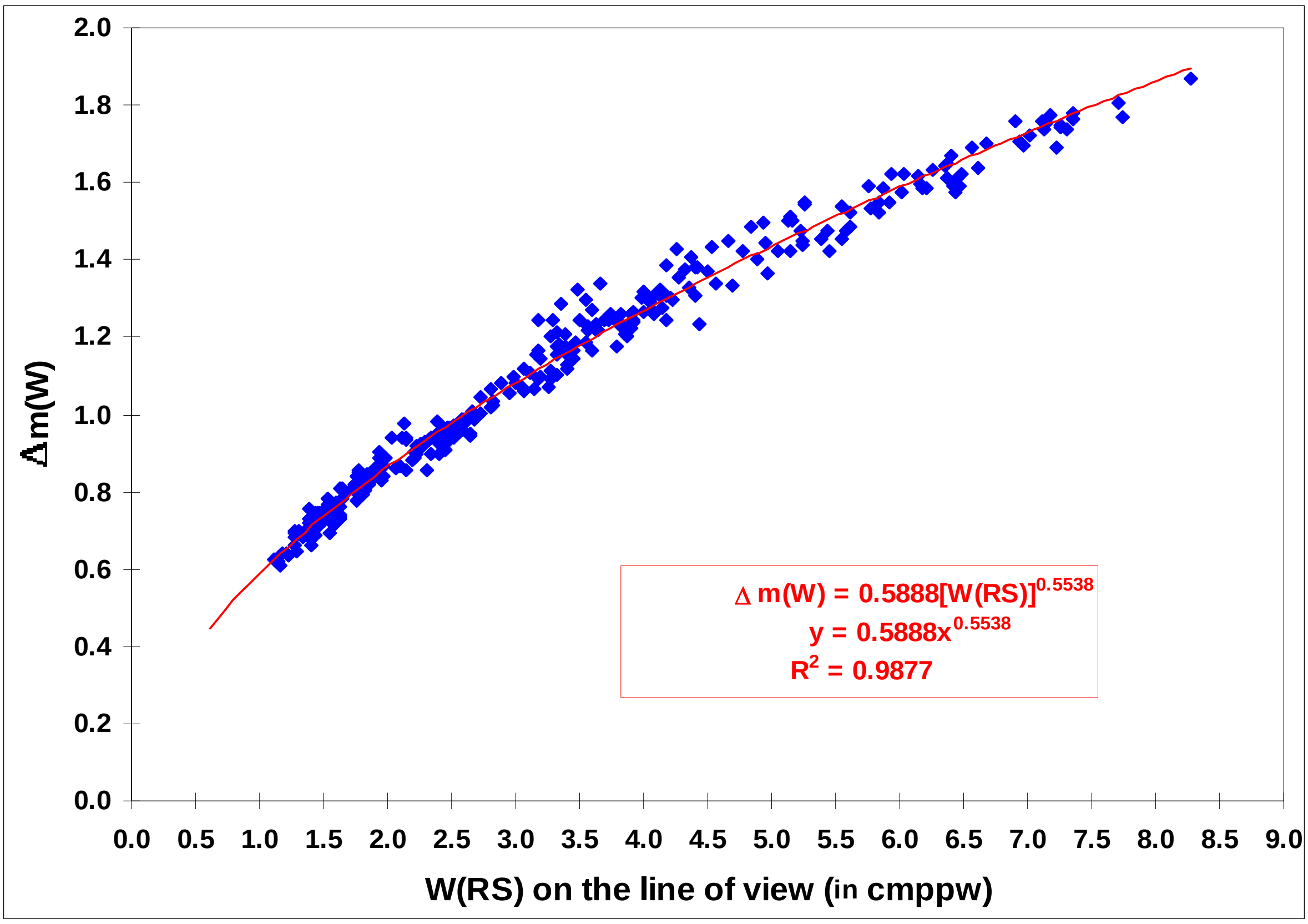

**Figure 3.** Approximation of the water vapour absorption magnitude Δm(W) by power function.

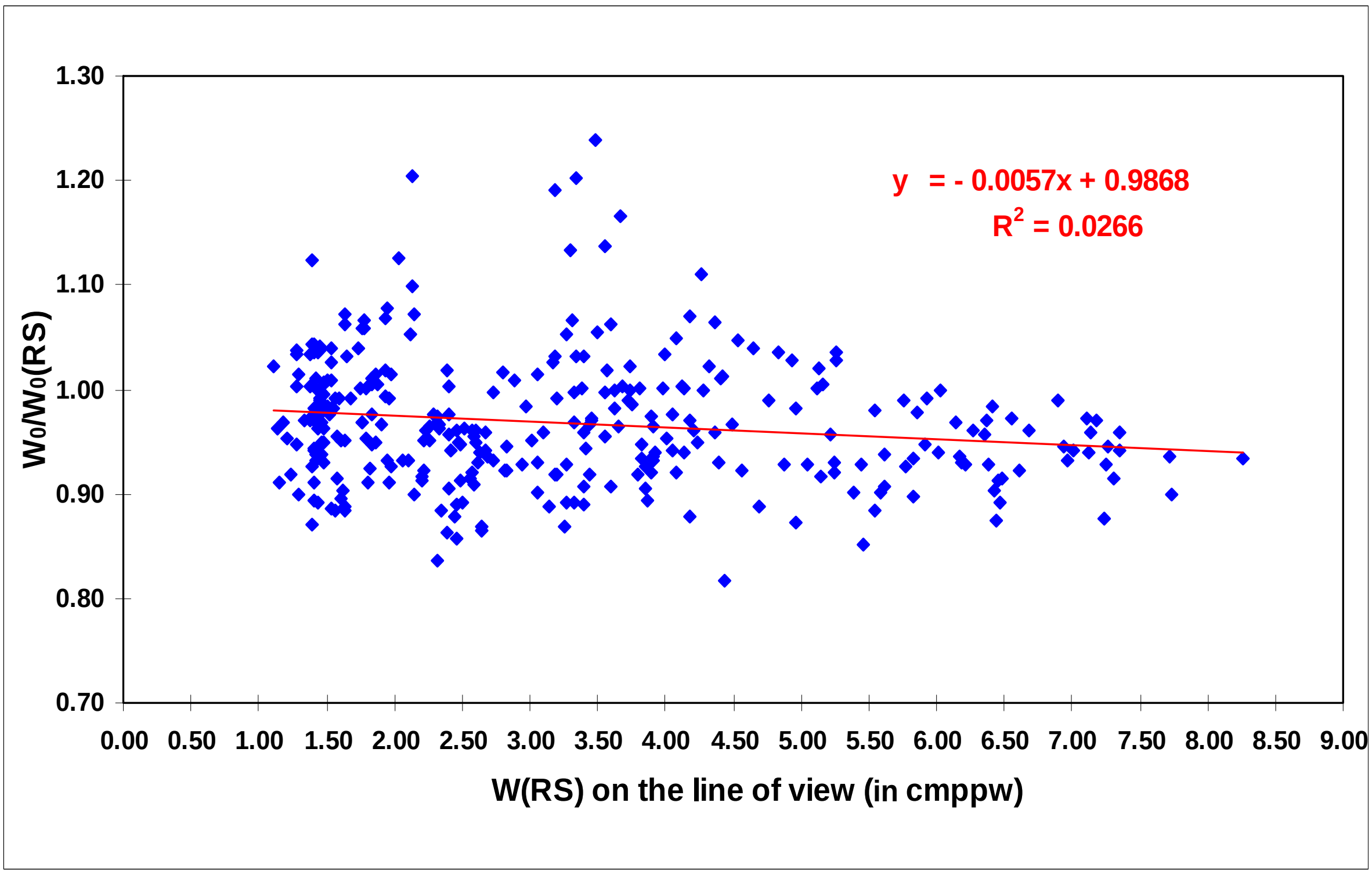

**Figure 4.** Comparison of $W_0$ (derived from starphotometer) with $W_0$(RS) (derived from interpolated radiosonde data), 09$^{th}$ to 14$^{th}$ of September 1999.

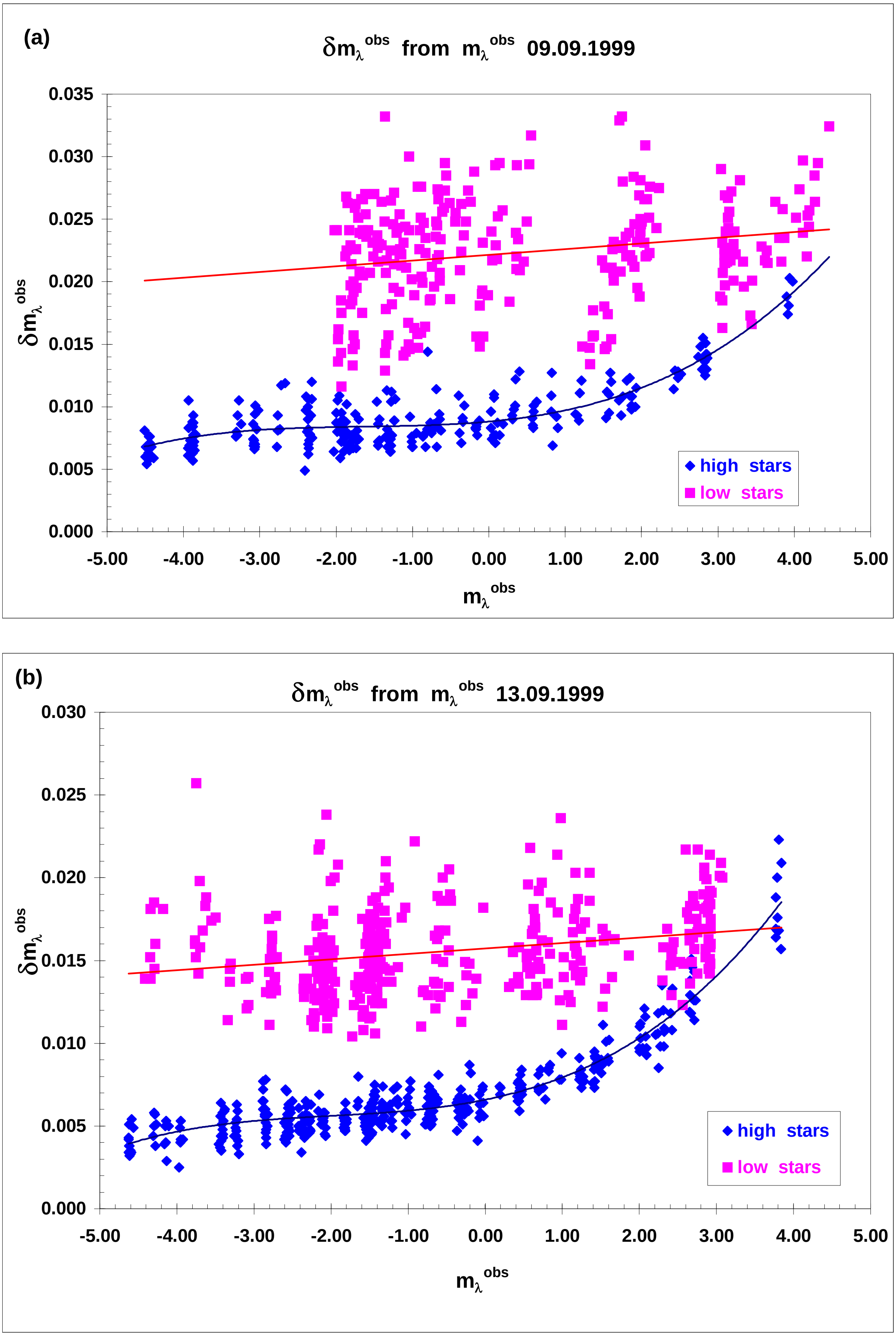

**Figure 5:** Errors $\delta m_\lambda^{obs}$ of the observed star magnitude $m_\lambda^{obs}$ in dependence on star magnitude.

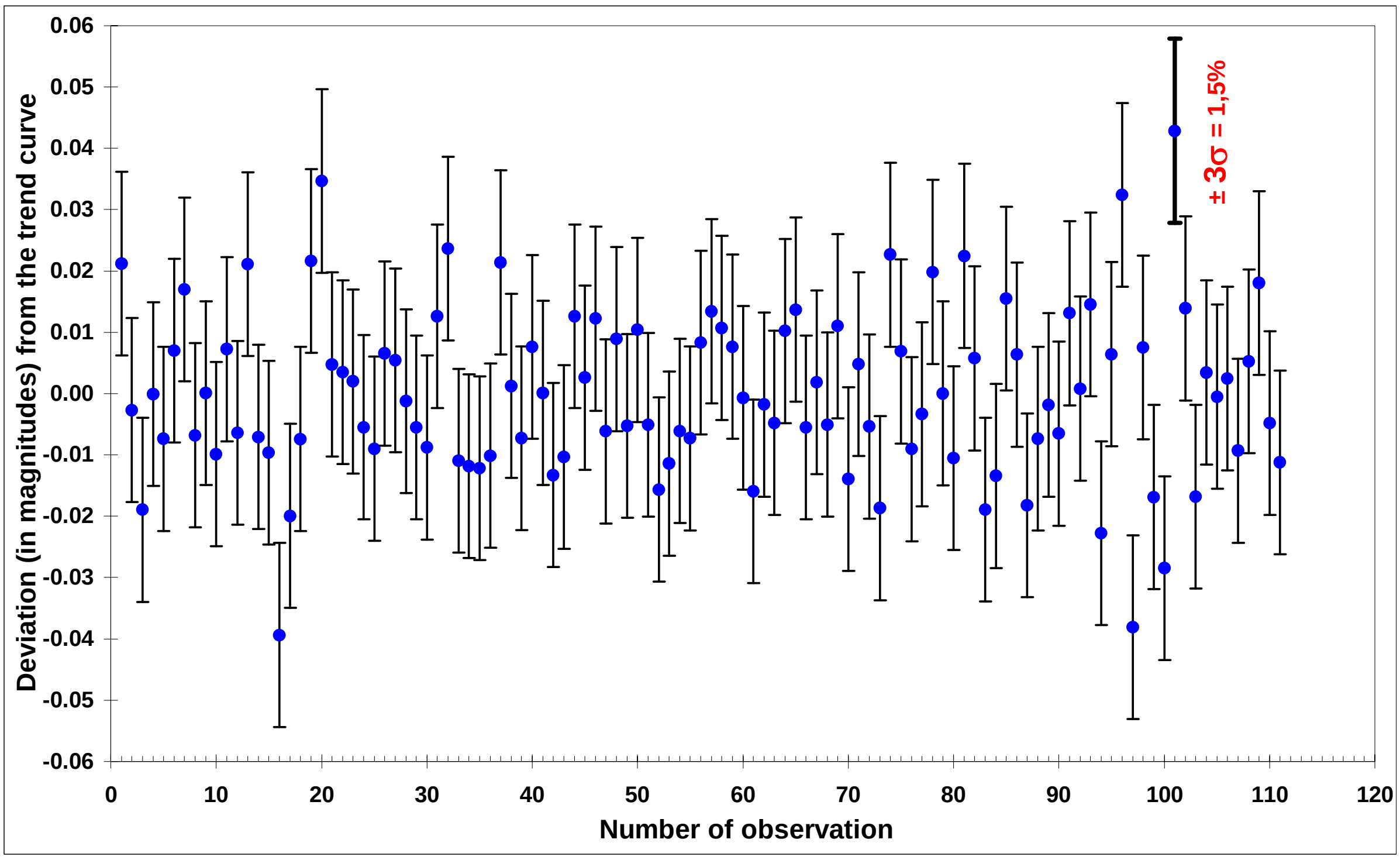

**Figure 6:** Deviations of single high star measurements (111 observations) to the mean trend curve of all 111 measurements in course of about 3.5 hours.

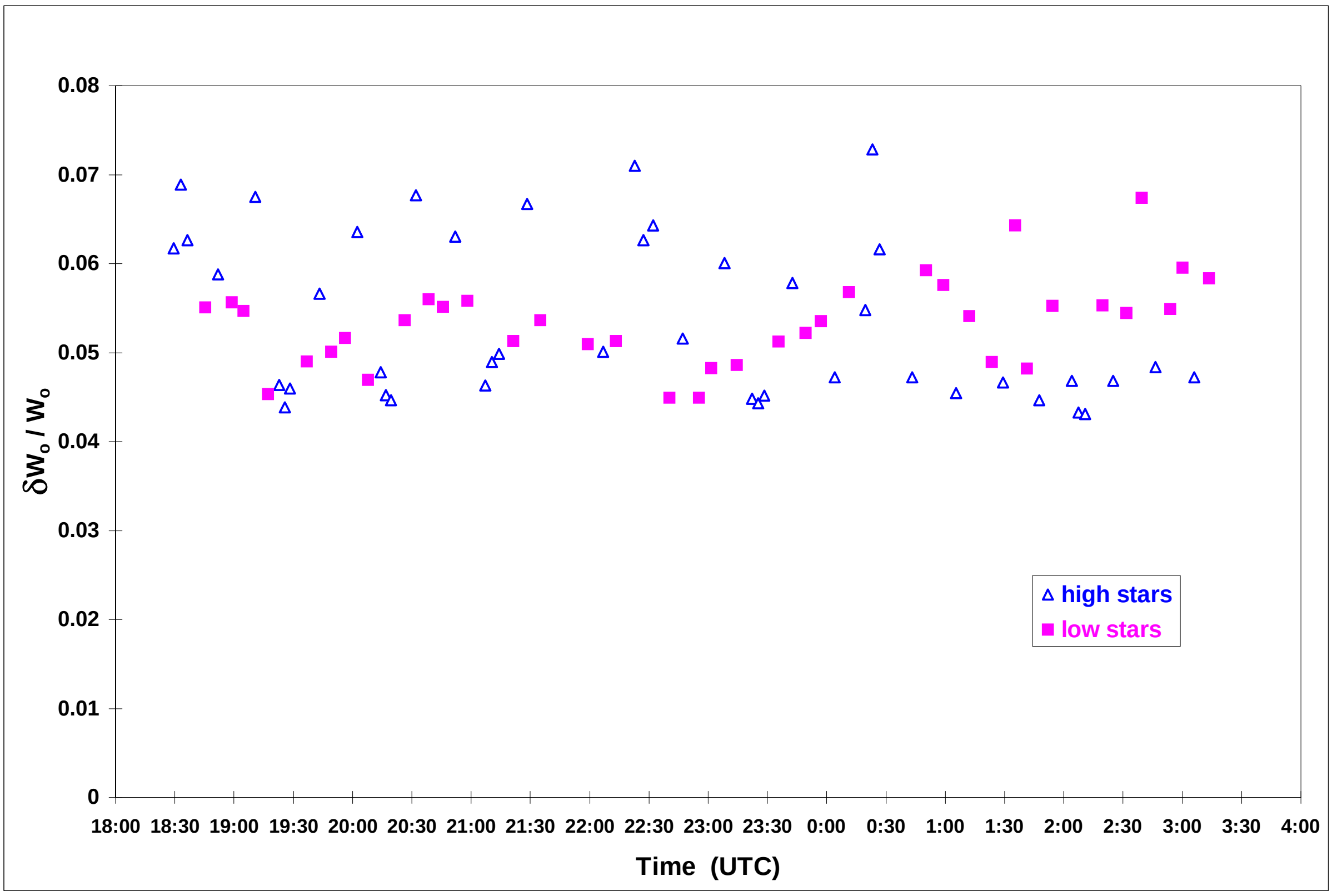

**Figure 8:** Relative error $\delta W_0/W_0$ of the water vapour content determination (starphotometer), 13th September 1999.

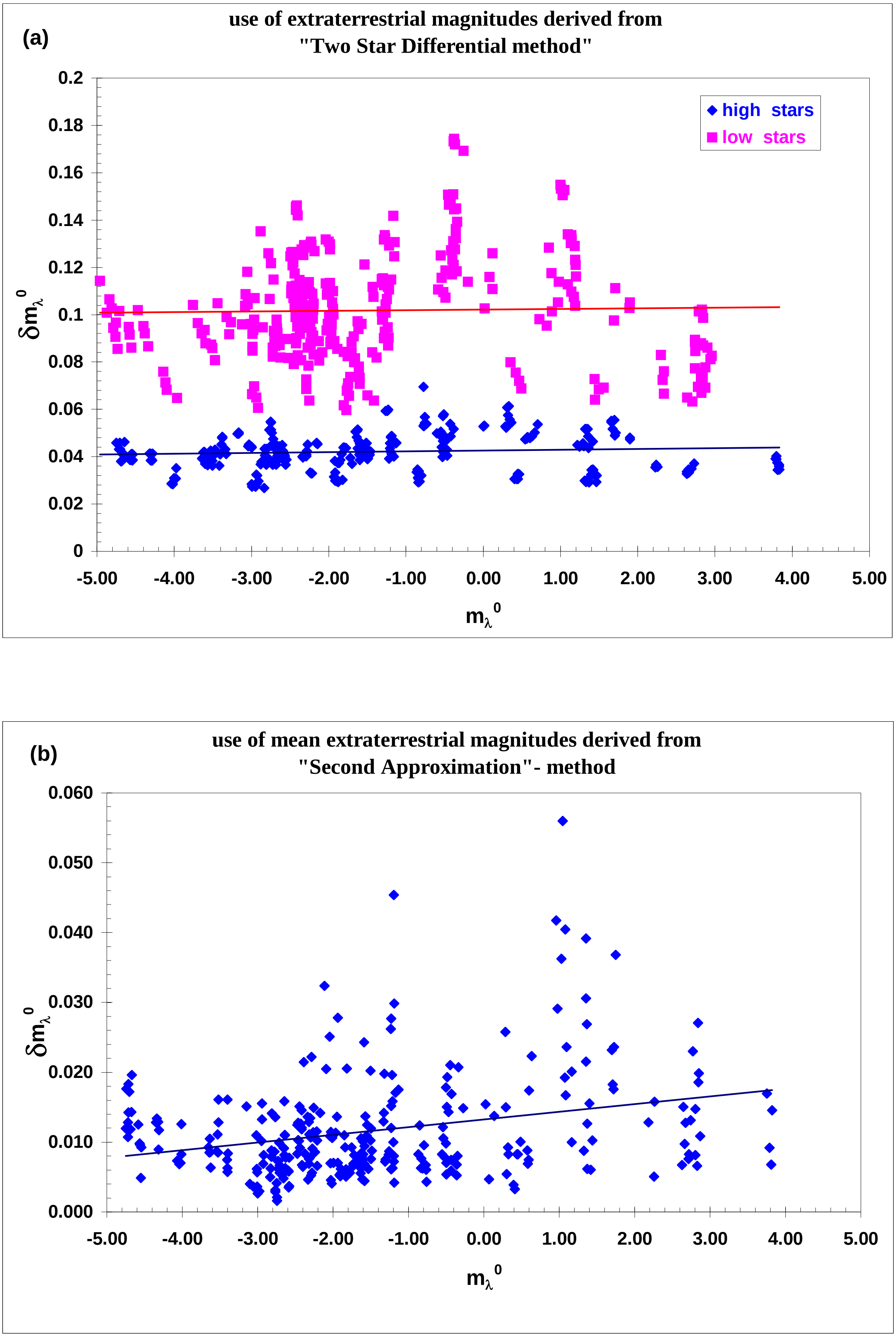

**Figure 7:** Errors $\delta m_\lambda^0$ in dependence on the extraterrestrial star magnitude $m_\lambda^0$.

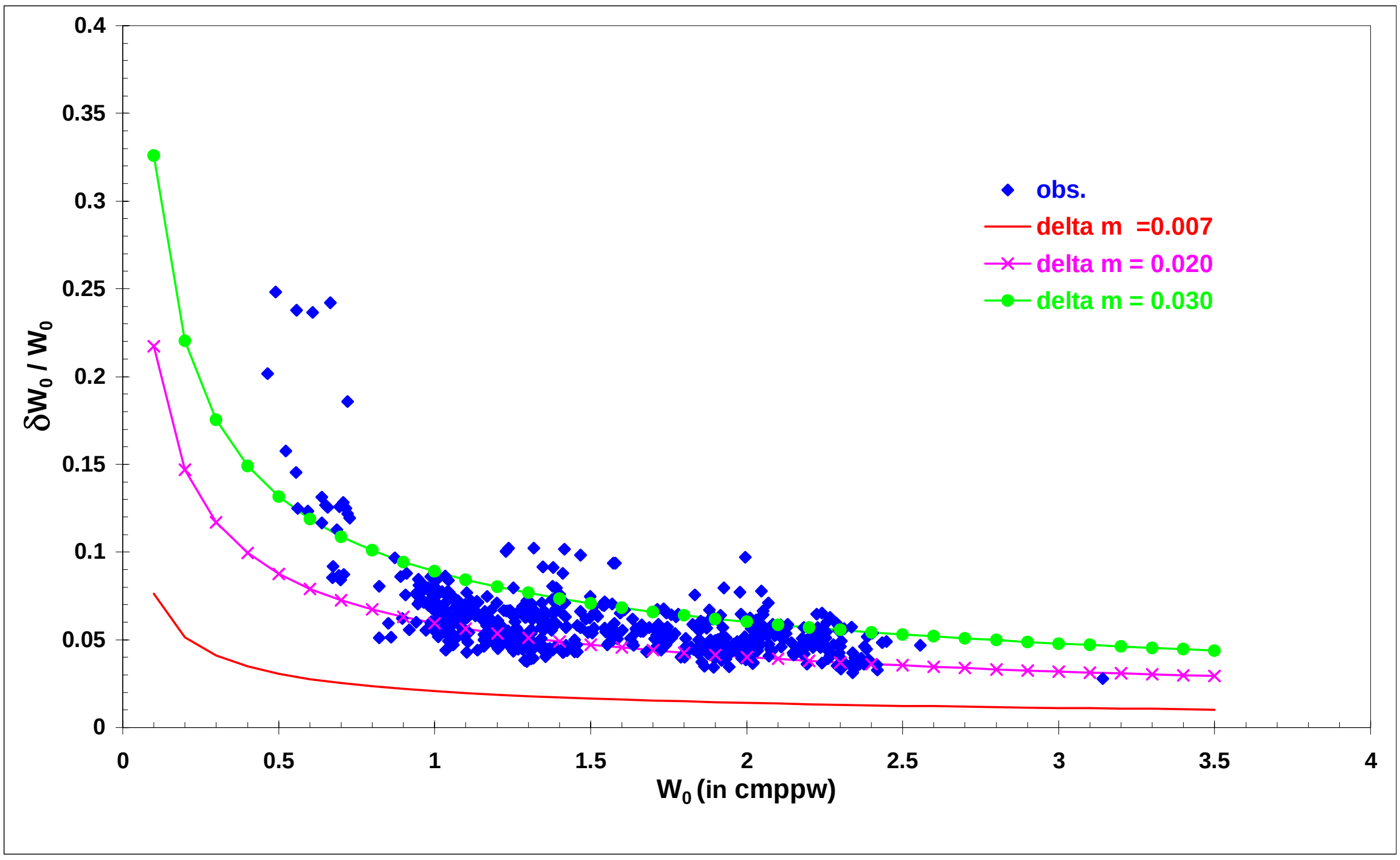

**Figure 9:** Relative error $\delta W_o/W_o$ (starphotometer) in dependence on water vapour column content $W_o$.

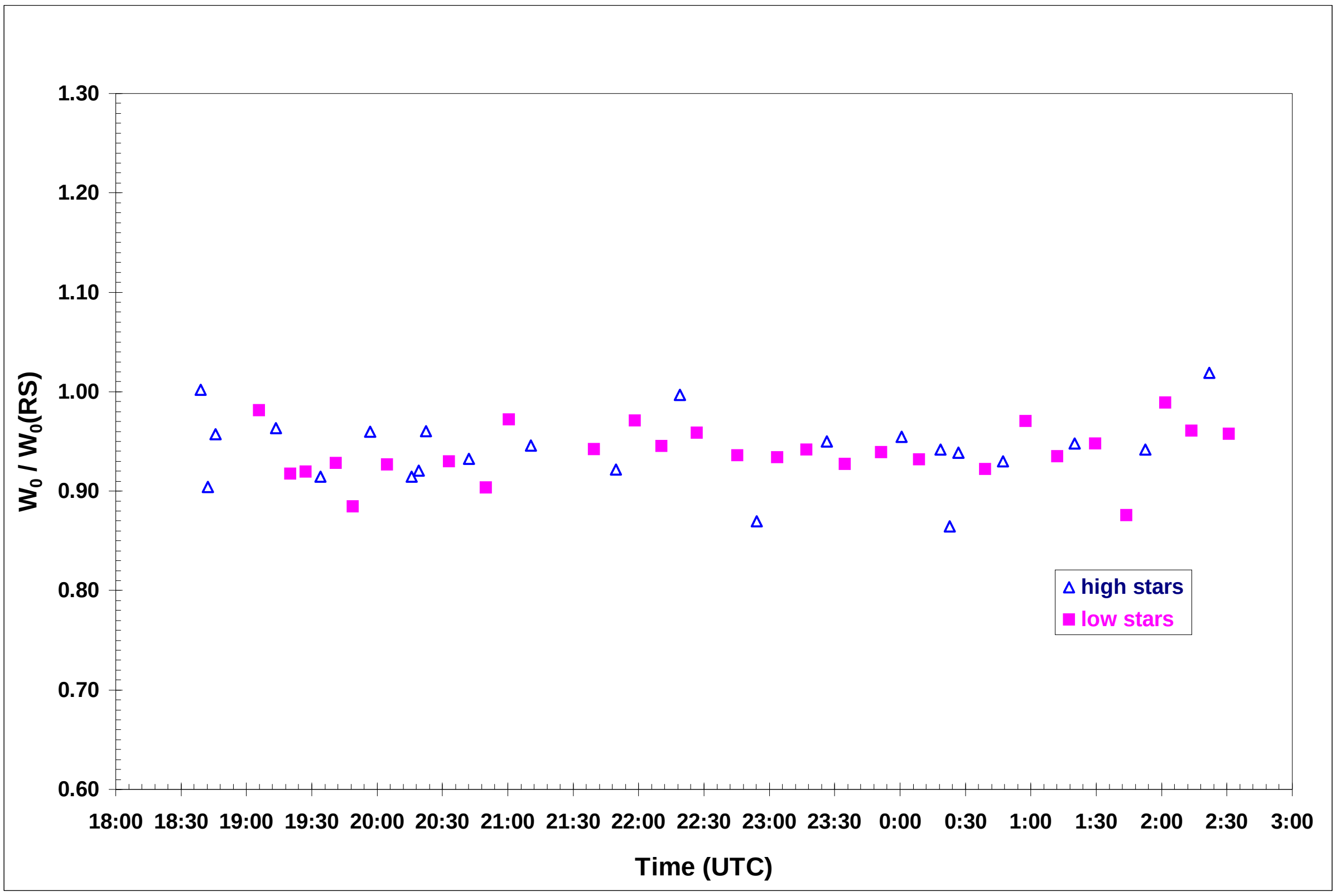

**Figure 10:** Comparison of water vapour columns derived from starphotometer ($W_0$) and interpolated radiosonde.

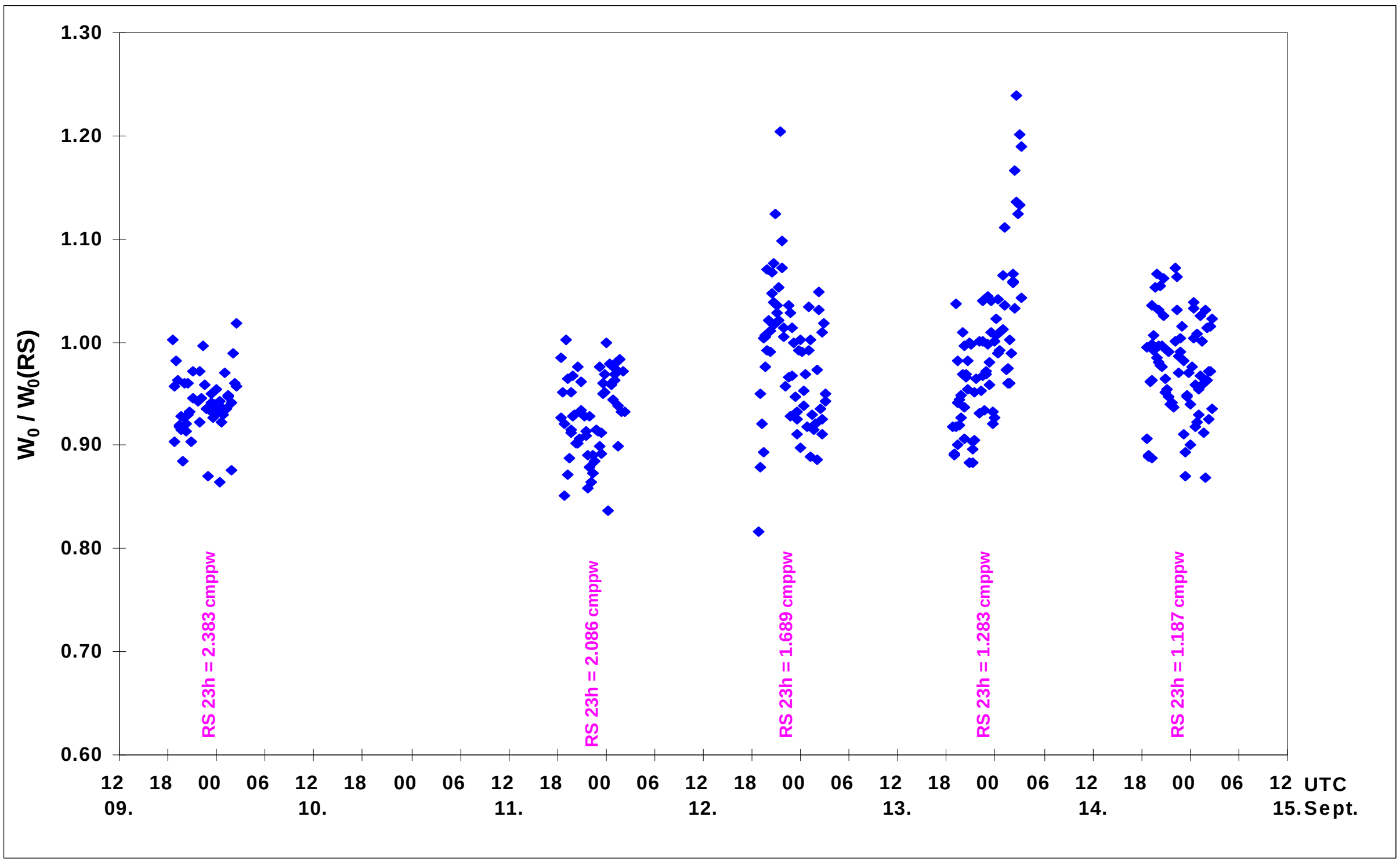

**Figure 11:** Comparison of $W_0$ and RS-data (interpolated) in the period 09th to 14th of September 1999.

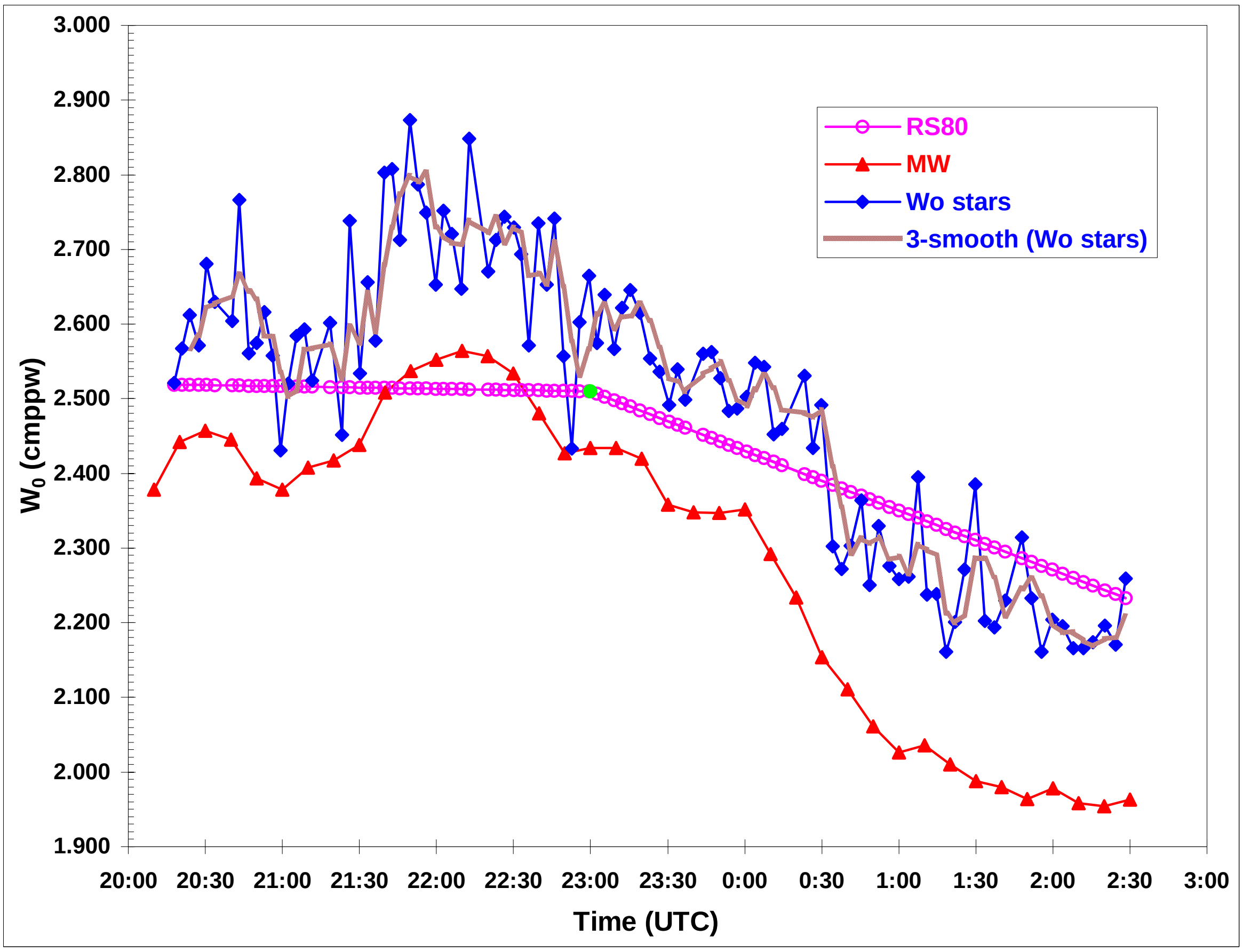

**Figure 12:** Rapid monitoring of column percipitable water $W_0$ by starphotometer, microwave-radiometer and radiosonde RS80 (interpolated). LACE experiment, August 11/12th 98.

## Appendix A

### Reconstruction of starphotometer MOL in 1999.

The original starphotometer MOL was built in 1993 (LEITERER et al., 1995) on the base of a SPCM-receiver (the avalanche silicon photo diode), and had an essential lack, namely: the micro-objective used in it does not provide a complete projection of the entrance pupil of the telescope to the very small sensitive platform (0.15 mm in diameter) of the receiver. So the receiver worked in a slightly divergent beam and used only a part of the emitted by the object light. Thus the requirements to guidance system of the telescope raised, and that lead to greater errors in the final results.

Therefore it was necessary to create a new optical system, which would build the image of exit pupil of the telescope on the platform with a diameter of about 0.1 mm. As in the considered case the exit pupil of telescope is located at a distance of 612 mm in front of the diaphragm and has a diameter of 60.2 mm, the system required should ensure a reduction of 602 times.

Such an optical system was developed in Pulkovo Observatory (ALEKSEEVA et al., 1995), and a new optical part was made for the Lindenberg-starphotometer.

This new optical part was mounted into the starphotometer in Lindenberg in June 1999. The initial investigations of the new variant of the photometer (including the investigation of “dead time“, see **Appendix B**) were made at the same time. In autumn of 1999 the regular observations were started.

The new optical construction is about twice as sensitive as the old one (see **Figure A1**) and corrects the bad optical situation for the 0.39 and 0.44 µm filters.

The observations are more accurate with new optical system how it can seen from **Figure A2a** and **Figure A2b**. So the error of observable star magnitude $\delta m_{\lambda}^{obs}$ is reduced for bright stars (e.g. $m_{obs}$ = -3.0) from 0.010 (1998 June 8th) to 0.08 (1999 September 9th).

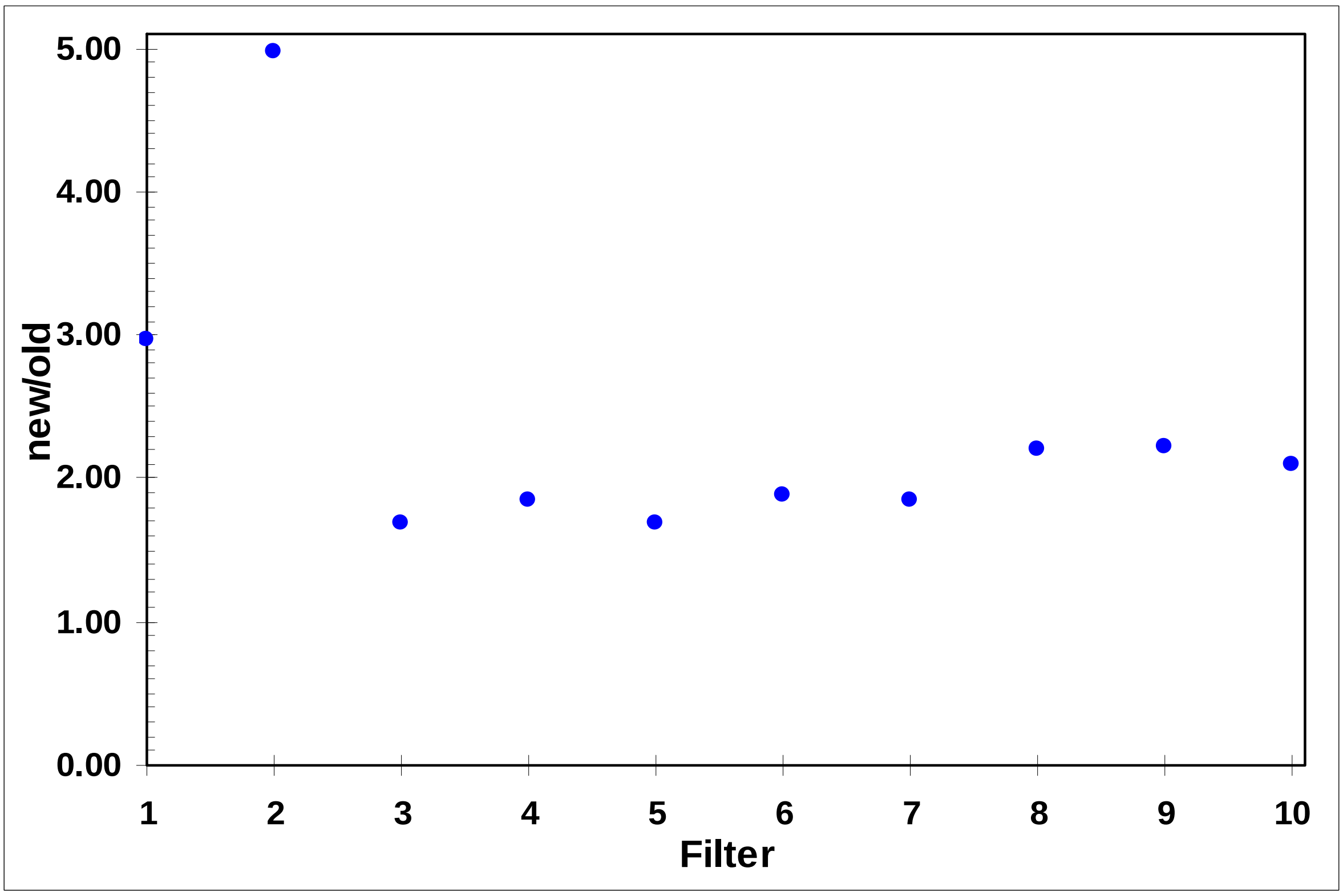

**Figure A1:** Gain of sensitivity after reconstruction of the starphotometer MOL in 1999.

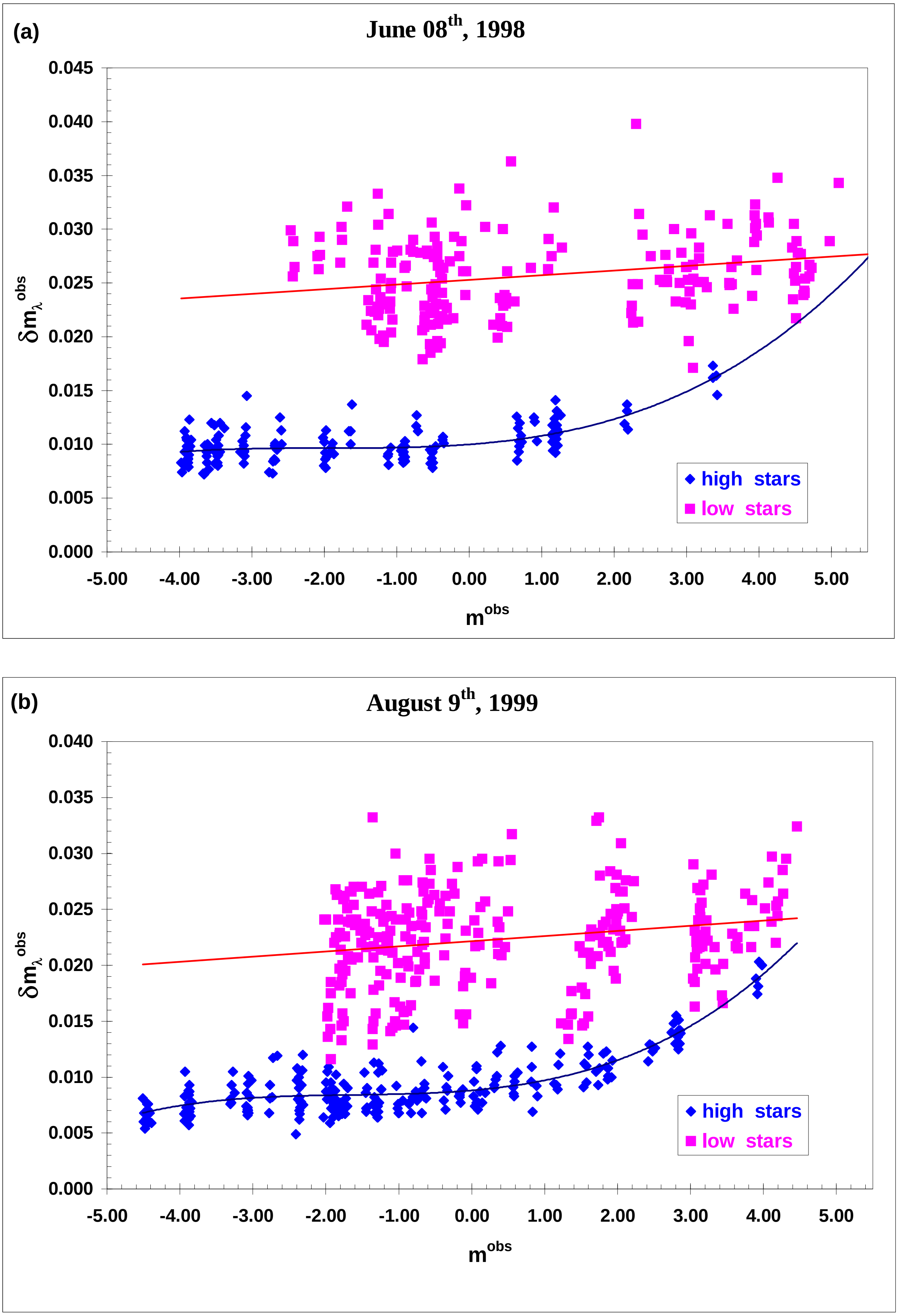

**Figure A2: Decreasing of observed star magnitude errors $\delta m_{\lambda}^{obs.}$ in dependence on the star magnitude $m^{obs.}$. Figure A2a before and Figure A2b after reconstruction of the optical part of the starphotometer.**

## Appendix B

### Dead Time investigation for SPCM-receiver.

### 1. Initial expressions and formulae.

The result count for SPCM-receiver may be approximated by the formula (according to Dr. R. Kalytis private communication):

$$\mathbf{U = U_0 \cdot e^{-U_0 \cdot \tau}} \qquad \textbf{(B1)},$$

where $\mathbf{U_0}$ - initial count (in count/sec); **U** - result count, registered by the SPCM-counter; $\tau$ - the so-called "extended"-type dead time (in sec). The graphs of this function are represented in **Figure B1** and **Figure B1a** for $\tau = 2{,}25 \cdot 10^{-7}$ sec (for Pulkovo's SPCM-starphotometer).

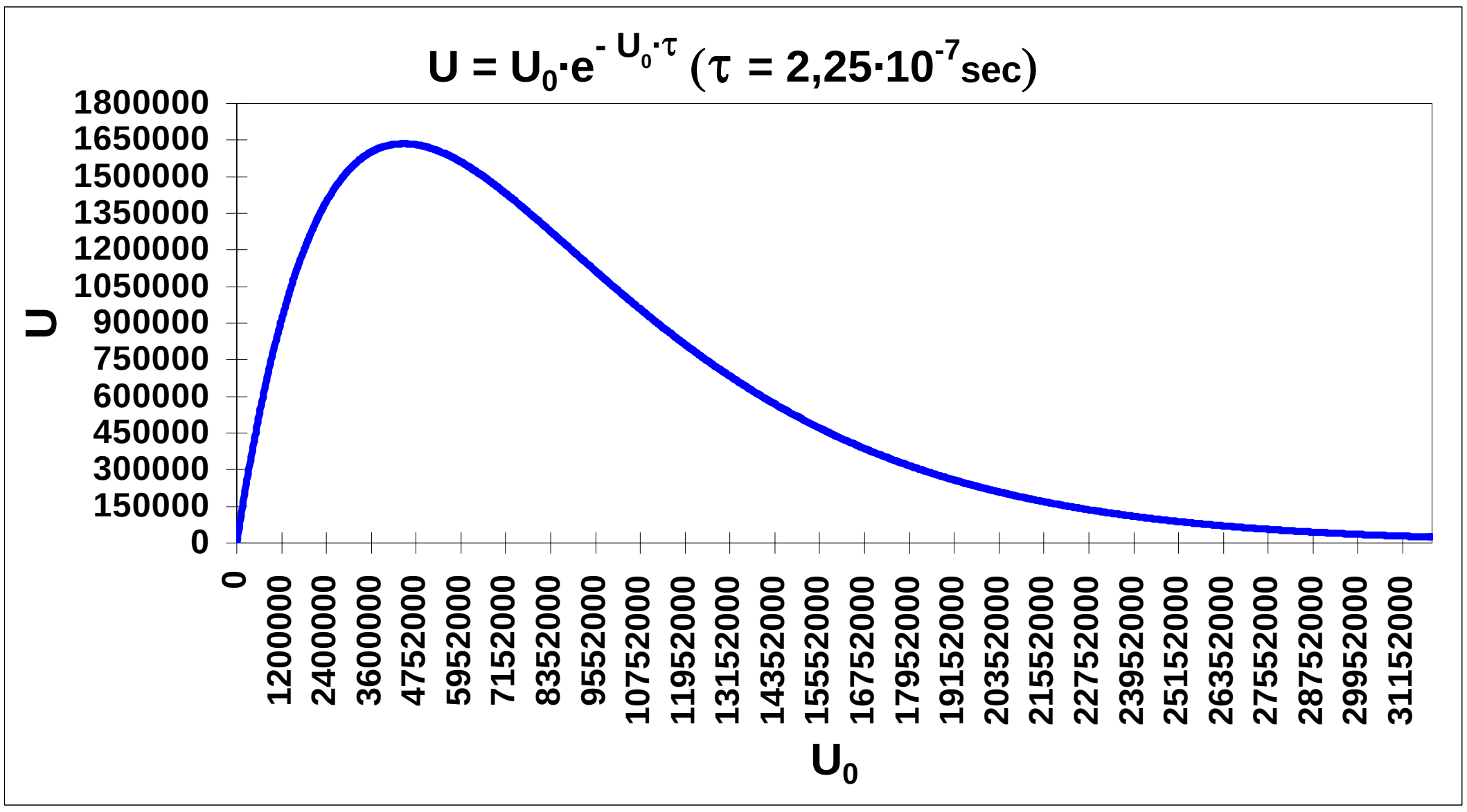

**Figure B1.**

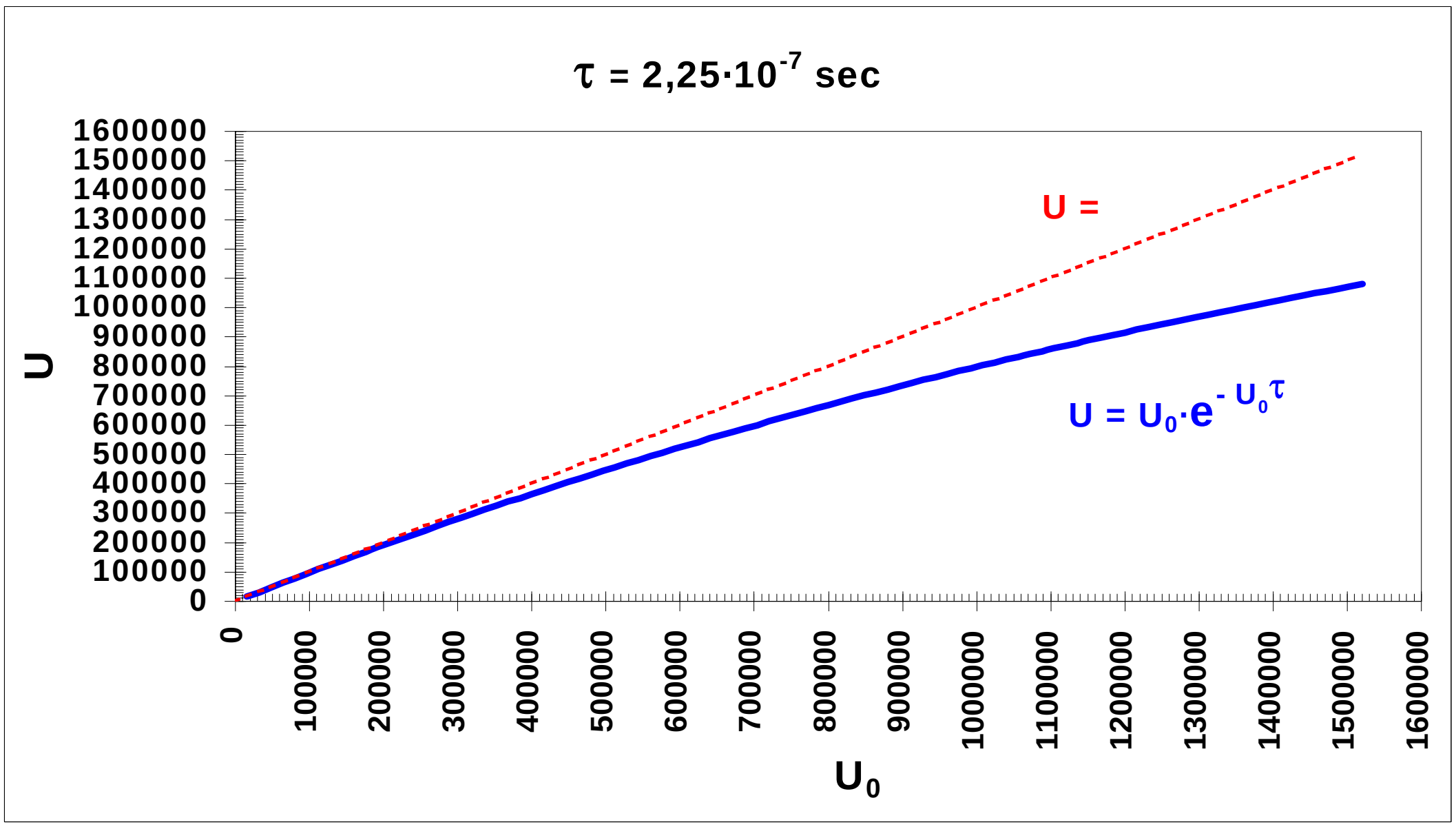

**Figure B1a.**

In practice it is necessary to know the inverse function $\mathbf{U_0 = f(U)}$ in order to correct registered counts to the dead time influence. For Function **(B1)** it is impossible to represent its inverse function in analytic form. But this problem may be solved using standard expansions:

$$e^{-x} = 1 - x/1! + x^2/2! - x^3/3! + x^4/4! - x^5/5! + x^6/6! - x^7/7! + ... \quad \mathbf{(B2)}$$

$$y = ax + bx^2 + cx^3 + dx^4 + ex^5 + fx^6 + gx^7 + ... \quad \mathbf{(B3)}$$

Inverse expansion for **(B3)**:

with $x = Ay + By^2 + Cy^3 + Dy^4 + Ey^5 + Fy^6 + Gy^7 + ...$ **(B3a)**,

$A = 1/a$; $B = - b/a^3$; $C = (2b^2 - ac)/a^5$; $D = (5abc - a^2d - 5b^3)/a^7$;

$E = (6a^2bd + 3a^2c^2 + 14b^4 - a^3e - 21ab^2c)/a^9$;

$F = (7a^3be + 7a^3cd + 84ab^3c - a^4f - 28a^2b^2d - 28a^2bc^2 - 42b^5)/a^{11}$;

$G = (8a^4bf + 8a^4ce + 4a^4d^2 + 120a^2b^3d + 180a^2b^2c^2 + 132b^6 - a^5g - 36a^3b^2e -$
$- 72a^3bcd - 12a^3c^3 - 330ab^4c)/a^{13}$

For Function **(B1)** the results are shown in the graphs of **Figure B2** and **Figure B3**:

$$U = U_0 - \tau \cdot U_0^2 + (0{,}5\tau^2) \cdot U_0^3 - (\tau^3/6) \cdot U_0^4 + (\tau^4/24) \cdot U_0^5 - (\tau^5/120) \cdot U_0^6 + (\tau^6/720) \cdot U_0^7 - ... \mathbf{(B4)}$$

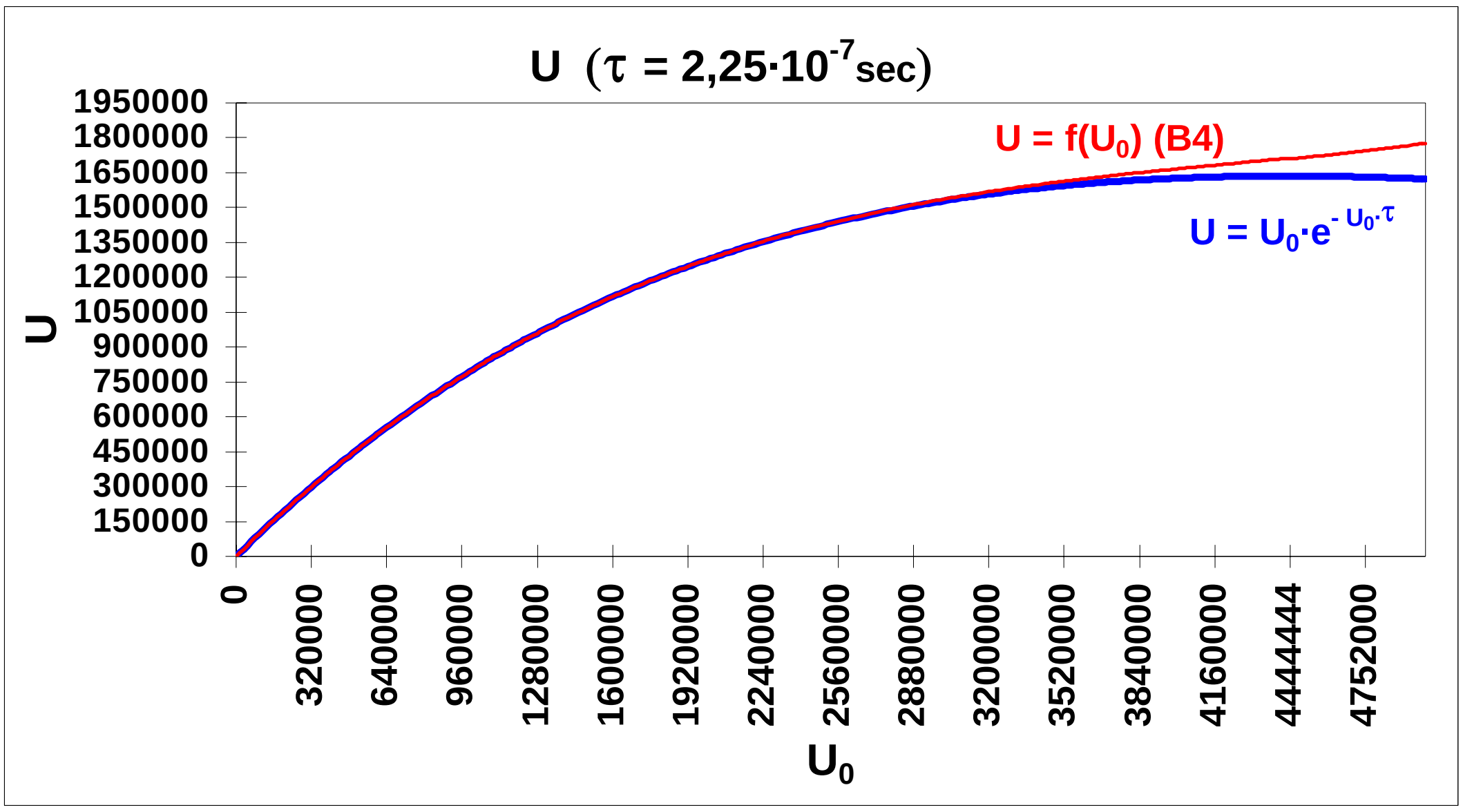

**Figure B2.**

$$U_0 = U + \tau \cdot U^2 + (1{,}5\tau^2) \cdot U^3 + (\tau^3 \cdot 8/3) \cdot U^4 + (\tau^4 \cdot 125/24) \cdot U^5 + (10{,}8 \cdot \tau^5) \cdot U^6 + (\tau^6 \cdot 16807/720) \cdot U^7 + .. \text{(B5)}$$

## 2. The practical determination of the so-called "extended"-type dead time.

In order to determine the real value of the so-called "extended"-type dead time $\tau$ we suggest the following method. The Function **(B1)** (see **Figure B1**) has only one maximum ($\mathbf{U_{max.}}$). Let us determine $\mathbf{U_{max.}}$ by means of the derivation of the Function **(B1)**:

$$\mathbf{U' = (U_0 \cdot e^{-U_0 \cdot \tau})' = U_0 \cdot (e^{-U_0 \cdot \tau})' + 1 \cdot e^{-U_0 \cdot \tau} = - U_0 \cdot \tau \cdot e^{-U_0 \cdot \tau} + \cdot e^{-U_0 \cdot \tau} = (1 - U_0 \cdot \tau) \cdot e^{-U_0 \cdot \tau}}$$

The Function **(B1)** has the maximum when **U'=0,** i.e.:

$$\mathbf{(1 - U_0 \cdot \tau) \cdot e^{-U_0 \cdot \tau} = 0;\ (1 - U_0 \cdot \tau) = 0;\ U_0 = \tau^{-1}.}$$

then

$$\mathbf{U_{max.} = U(\tau^{-1}) = \tau^{-1} \cdot e^{-1}}$$

and finally:

$$\mathbf{\tau = (U_{max.} \cdot e)^{-1}} \quad \mathbf{(B6)}$$

Therefore it is sufficient to determine with high accuracy the value $U_{max.}$ from laboratory measurements (for example with calibrated band-lamp source with adjusting voltage level) in order to know the real value of dead time $\tau$ for specimen exemplar of SPCM-receiver according to Equation **(B6)**.

The correction of registered counts to dead time influence may be reduced according to expansion **(B5)**. This expansion has the sufficient accuracy up to U ~ 1200000 - 1300000 count/sec for $\tau = 2{,}25 \cdot 10^{-7}$ sec (see **Figure B3**).

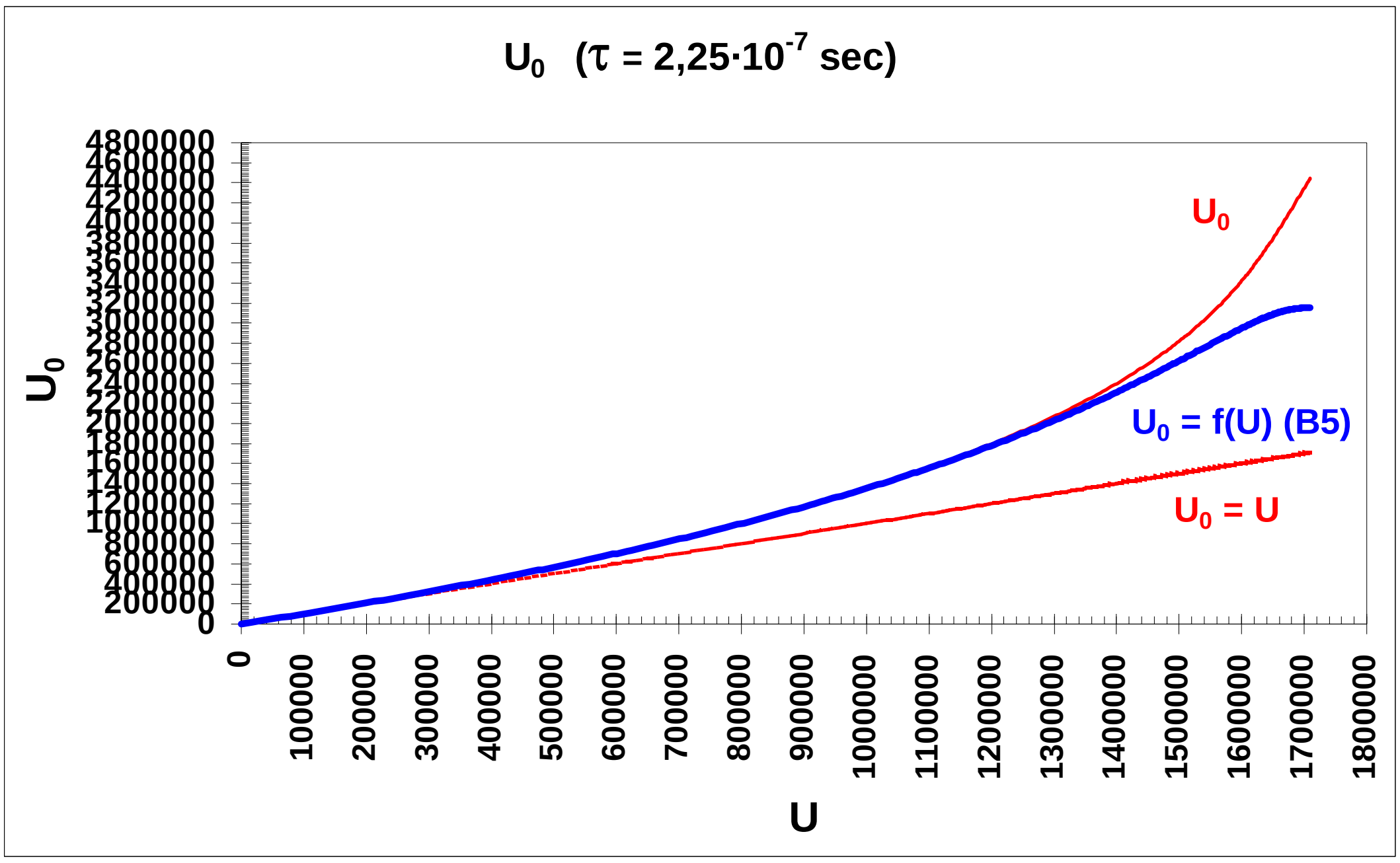

**Figure B3.**

**3. The comparison of two types of dead time ("extended" and "non-extended").**

All receivers may be divided into two groups according to types of dead time:

1) The receivers with so-called "extended"-type dead time (such as SPCM).
2) The receivers with so-called "non-extended"-type dead time (such as photomultipliers).

The receivers of **first** type (SPCM) was described above. Theirs parameters are represented by the Equations **(B1)**, **(B4), (B5), (B6)**.

The parameters of receivers of the **second** type (photomultipliers) may be approximated by the analytic equations :

$$U = U_0 \cdot (1 + U_0 \cdot \tau)^{-1} \qquad \textbf{(B7)}$$

$$U_0 = U \cdot (1 - U \cdot \tau)^{-1} \qquad \textbf{(B8)}$$

$$\tau = \frac{K \cdot U_1 - U_2}{U_1 \cdot U_2 \cdot (K-1)}; \; K = \frac{(U_0)_2}{(U_0)_1} \qquad \textbf{(B9)}$$

Here Equations **(B7)** and **(B8)** were taken from the paper by D.A. Ralys and R. Kalytis (Bull. Vilnius Obs. 1978, No. 48, p. 3-17), and formula **(B9)** was obtained by means of the following method:

We used the one laboratory calibrated band-lamp source and two area-calibrated input diaphragms for our starphotometer. Then the coefficient K will be:

$$K = \frac{(U_0)_2}{(U_0)_1} = \frac{S_2}{S_1},$$

where $S_1$ and $S_2$ - areas of input diaphragms 1 and 2 respectively. Then on the base of Formula (B**8**) for result the counts $U_1$ and $U_2$, registered by the receiver through input diaphragms 1 and 2 respectively, we obtained the expression (B9) for dead time $\tau$.

The graphic comparison for two types of receivers (according to Formulae **(B1)** and **(B7)**) is represented in **Figure B4** for $\tau = 2{,}25 \cdot 10^{-7}$ sec.

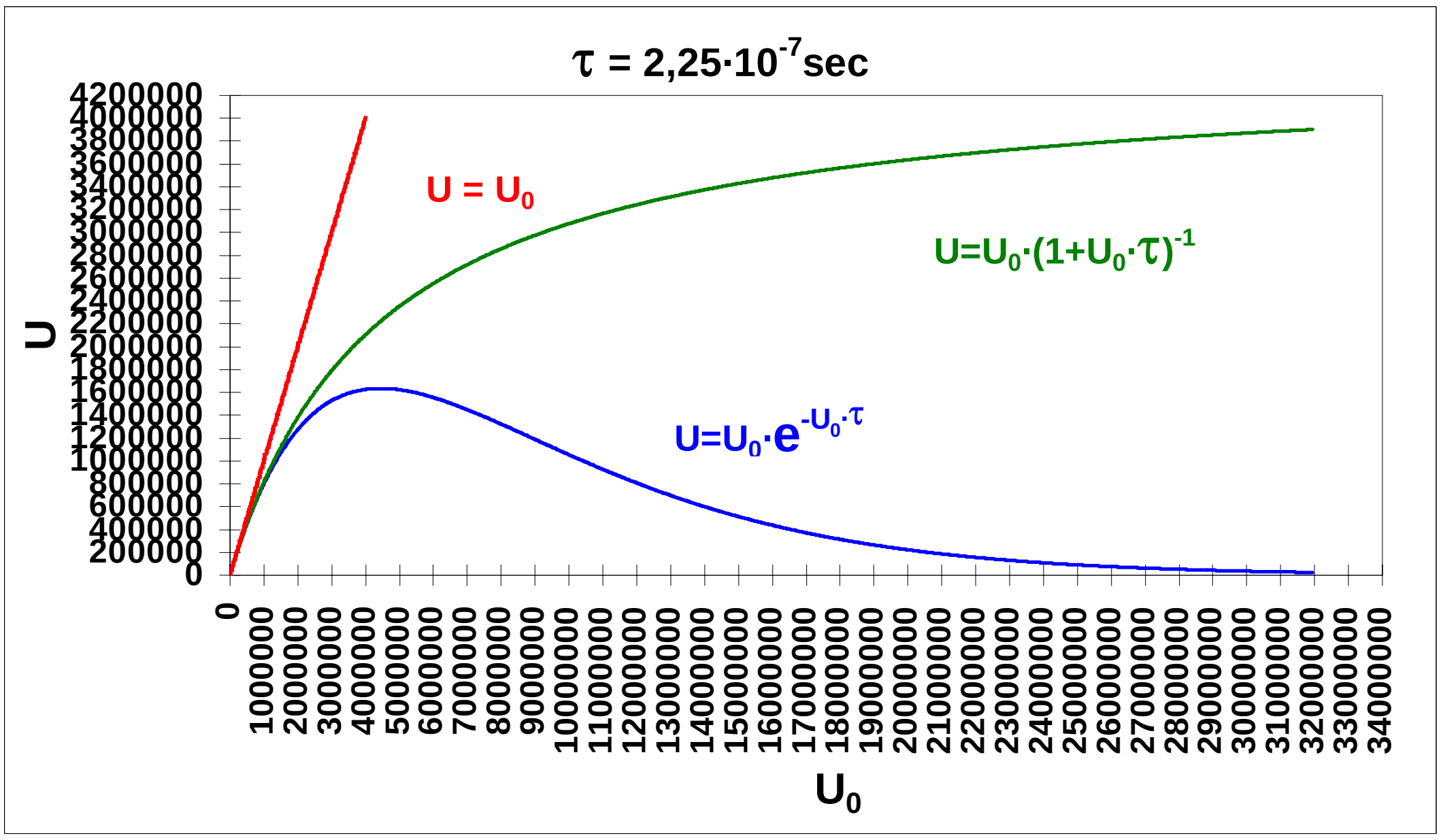

**Figure B4.**

Sometimes it happens in practice (for example for starphotometer MOL) that the dependence [Equation **(B8)**] and corresponding $\tau$ are used for dead time correction for receivers of the **first** type (SPCM). This procedure is incorrect. In order to confirm that we obtained the value of dead time $\tau_{\text{"2"}}$ = **(3,21 ± 0,47)·10⁻⁷ sec** for the Pulkovo SPCM-starphotometer on the base of Equation **(B9)** and the corresponding method described above. We would remind you that we had the real $\tau$ = **(2,25 ± 0,01)·10⁻⁷ sec** according our measurements on the base of Equation **(B6)**. The corresponding graphic comparisons for U and $U_0$ are represented in **Figure B5** and **Figure B6**.

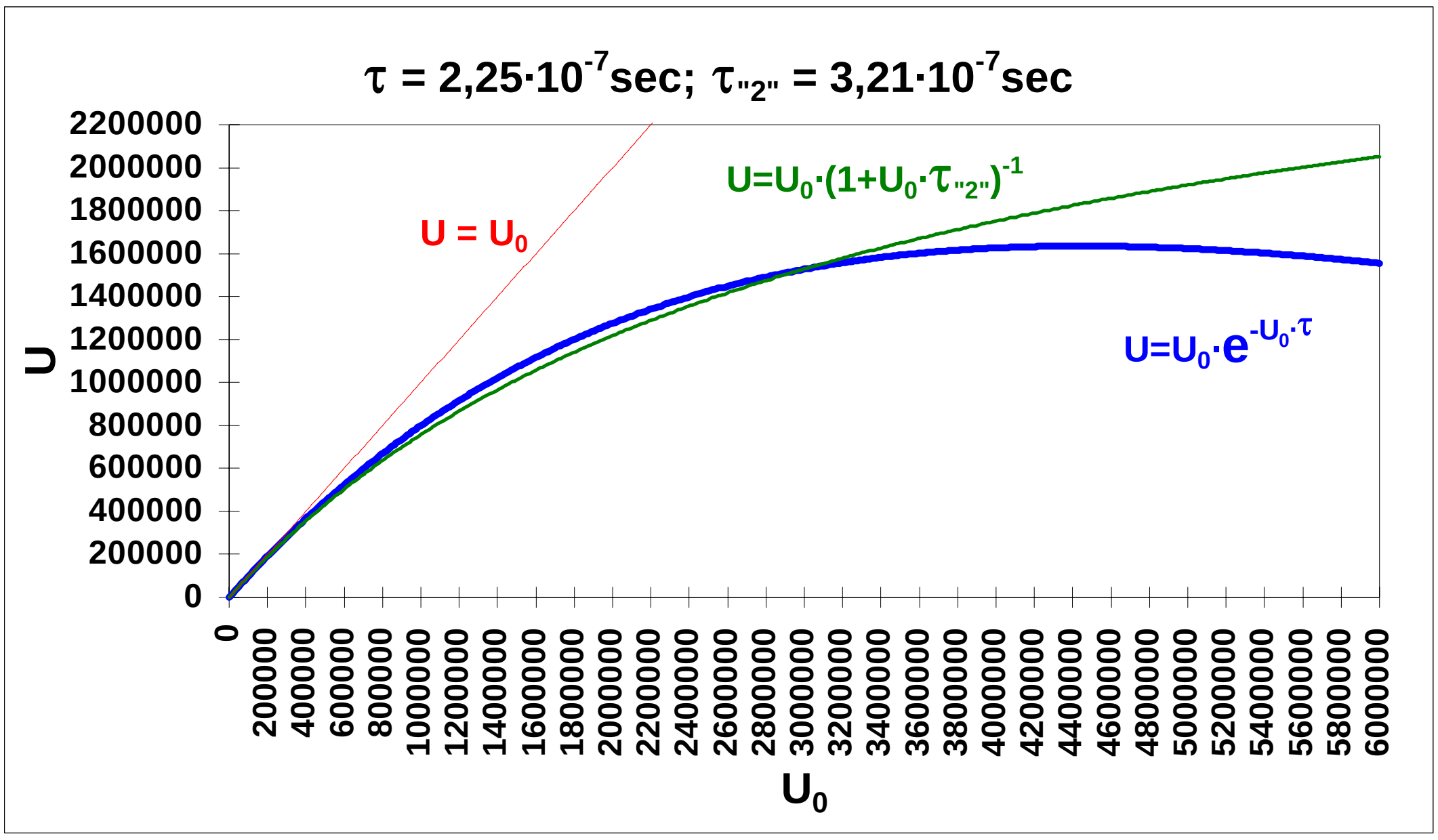

**Figure B5.**

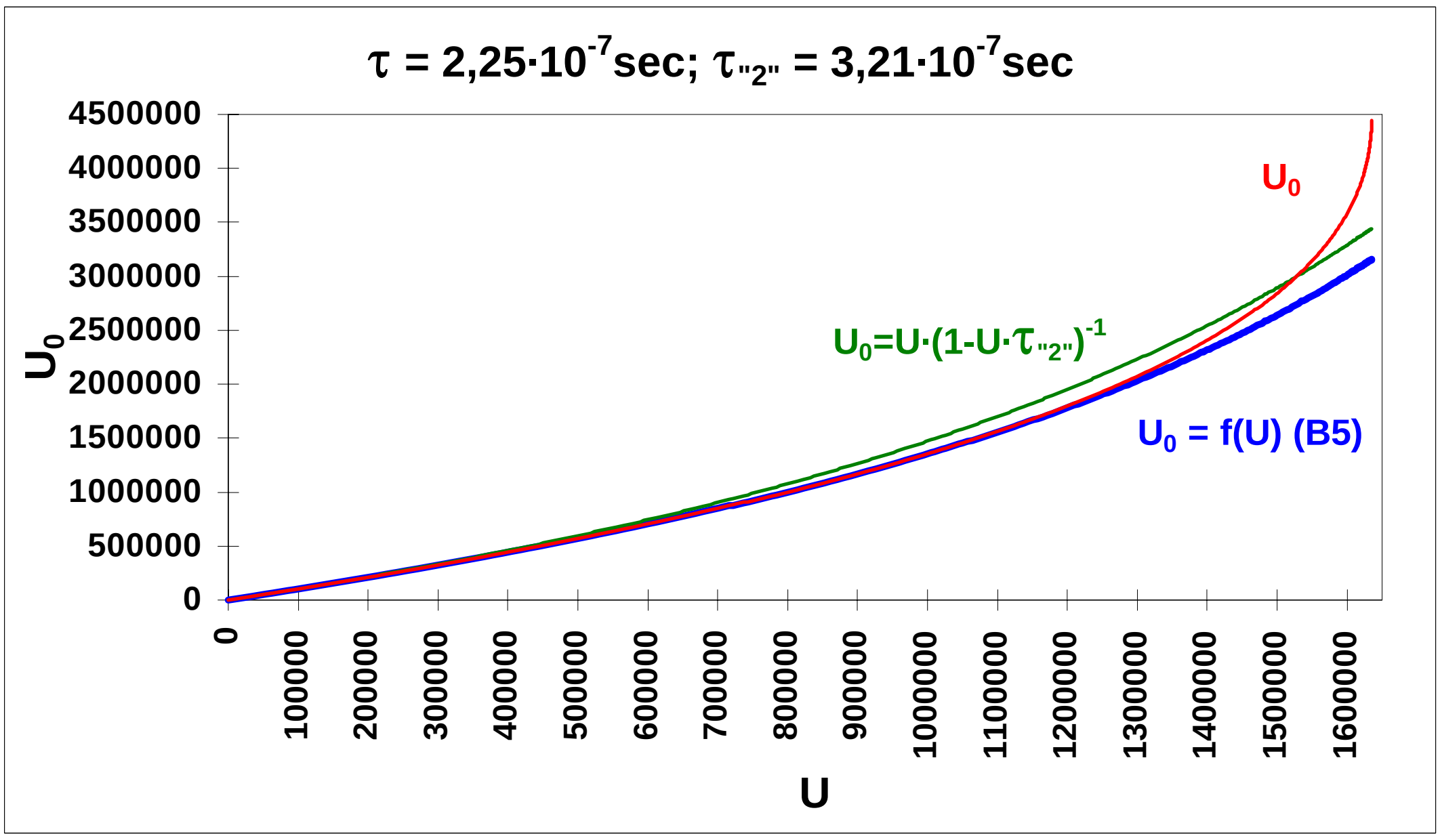

**Figure B6.**

As one can see the difference becomes appreciable already for U ~ **500000**.

## 4. Dependence $\tau$ (dead time) on temperature.

Our posterior investigations made in Pulkovo and Lindenberg have separately shown that there are dependencies of value $\tau$ on temperature. Those dependencies are different for different examples of SPCM-receivers (see **Figure B7** and **Figure B8**). In both cases $\tau$ decreases with temperature, but the speeds and characters of decreasing are different for Pulkovo's and Lindenberg's photometers. Especially the most careful investigation of $\tau$ - temperature dependence is needed for low (negative) temperatures where we do not have observed data yet.

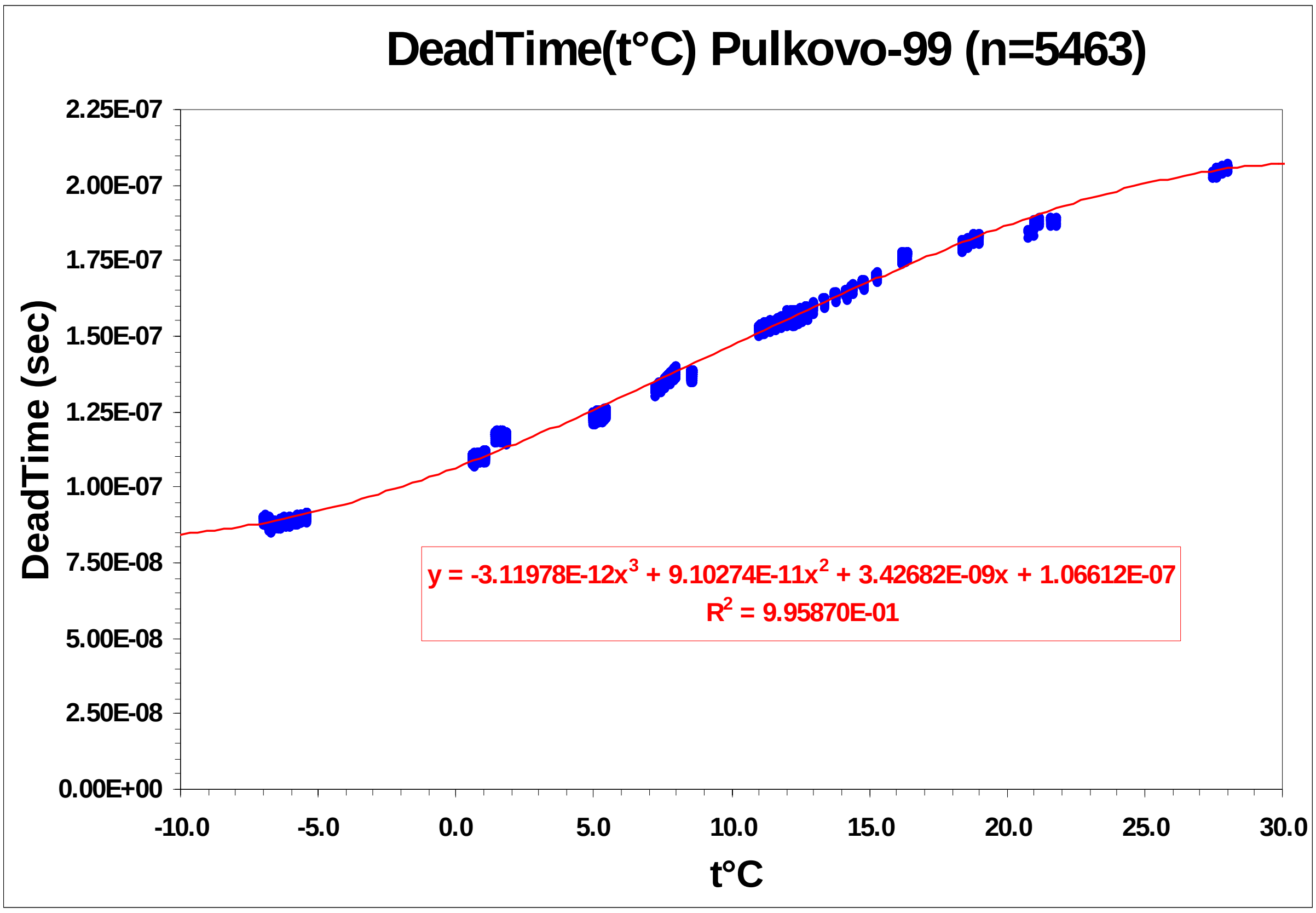

**Figure B7. Dead time dependence on temperature for Pulkovo's star photometer.**

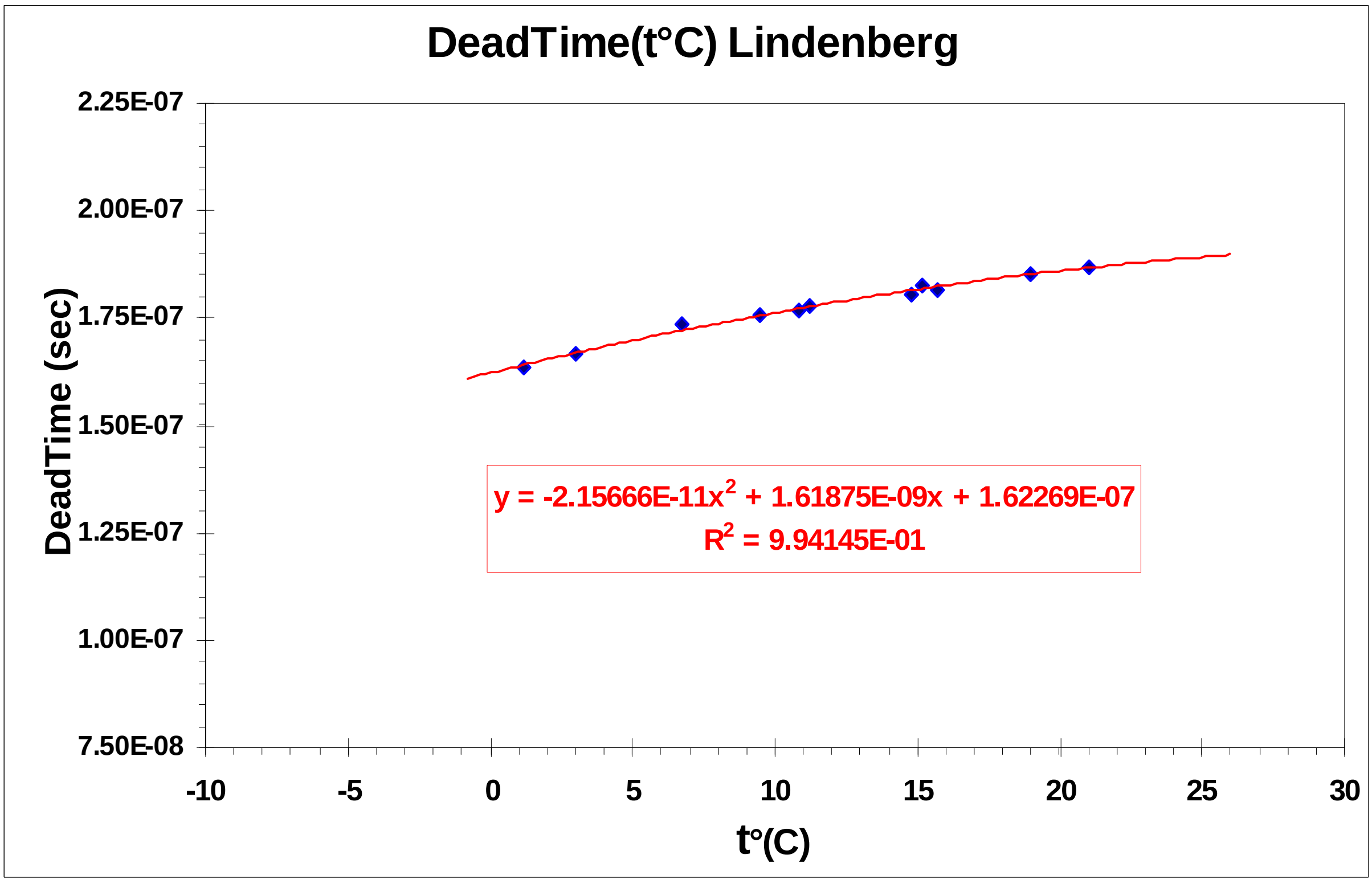

**Figure B8. Dead time dependence on temperature for Lindenberg's star photometer.**